\documentclass[a4paper,fleqn,usenatbib]{mnras}

\usepackage{upgreek}
\usepackage[T1]{fontenc}

\usepackage{graphicx}   
\usepackage{amsmath}    
\usepackage{amssymb}    

\newcommand{\fx}{\ifmmode \nu_{\rm x}\else$\nu_{\rm x}$\fi}
\newcommand{\Px}{\ifmmode P_{\rm x}\else$P_{\rm x}$\fi}
\newcommand{\fF}{\ifmmode \nu_{\rm 0}\else$\nu_{\rm 0}$\fi}
\newcommand{\fO}{\ifmmode \nu_{\rm 1}\else$\nu_{\rm 1}$\fi}
\newcommand{\fSO}{\ifmmode \nu_{\rm 0.61}\else$\nu_{\rm 0.61}$\fi}
\newcommand{\RRSO}{RR$_{0.61}$~}

\title[RR Lyrae stars in NGC~6362]{RR Lyrae stars in NGC~6362\thanks{Based on data obtained with  du Pont and Swope telescopes at Las Campanas Observatory.}}

\author[Smolec et al.]
{R. Smolec,$^{1}$
 P. Moskalik,$^{1}$\thanks{E-mail: pam@camk.edu.pl}
 J. Ka\l{}u\.zny,$^{1}$\thanks{deceased}
 W. Pych$^{1}$
 M. R\'o\.zyczka,$^{1}$\and
 and I. B. Thompson$^{2}$
\\
$^{1}$Nicolaus Copernicus Astronomical Center
              of the Polish Academy of Sciences,
              Bartycka 18, PL-00-716 Warsaw\\
$^{2}$The Observatories of the Carnegie Institution for Science,
              813 Santa Barbara Street, Pasadena, CA 91101, USA\\
}

\date{Accepted 2017 January 1. Received 2017 January 10; in original form 2016 October 11}

\pubyear{2016}

\begin{document}
\label{firstpage}
\pagerange{\pageref{firstpage}--\pageref{lastpage}}
\maketitle

\begin{abstract}
We present the analysis of the top-quality photometry of RR~Lyrae
stars in the globular cluster NGC~6362, gathered over 11 observing
seasons by the CASE project. 16 stars are fundamental mode pulsators
(RRab stars) and 16 are first overtone pulsators (RRc stars). In two
stars, previously identified as RRab, V3 and V34, we detect
additional periodicity identified as radial first overtone mode.
Lower than usual period ratios ($0.730$ and $0.728$), dominant
pulsation in the radial fundamental mode and presence of a
long-period modulation indicate, that these two variables are not
classical RRd stars, but are new members of the recently identified
class of anomalous RRd variables. In a significant fraction of RRc
stars, $63$\thinspace per cent, we detect additional shorter-period
variability in the $(0.60,\,0.65)P_1$ range. This form of
double-periodic pulsation must be common in first overtone RR~Lyr
stars, as space observations indicate. The incidence rate we find in
NGC~6362, is the highest in ground-based observations reported so
far. We study the properties of these stars in detail; in particular
we confirm that in the colour-magnitude diagram, this group is
adjacent to the interface between RRab and RRc stars, as first
reported in the analysis of M3 observations by Jurcsik et al. The
incidence rate of the Blazhko effect is also very high: we observe
it in $69$\thinspace per cent of RRab stars and in $19$\thinspace
per cent of RRc stars. Rare, double-periodic modulation is reported
in one RRab and in one RRc star. Finally we discuss V37 -- a
peculiar variable in which we detect two close high-amplitude
periodicities and modulation. Its previous classification as RRc
must be treated as tentative.
\end{abstract}




\section{Introduction}\label{sect:intro}

RR~Lyrae stars are low-mass, horizontal-branch, pulsating variables.
They are divided into three main classes: fundamental mode, radial
pulsators (RRab stars), first overtone, radial pulsators (RRc stars)
and double-mode pulsators, pulsating in the radial fundamental and
in the radial first overtone modes simultaneously (RRd stars). In
the last few years, the top-quality photometry of RR~Lyr stars
gathered by the space telescopes ({\it MOST}, {\it CoRoT} and {\it
Kepler}) and by the ground-based photometric sky surveys (e.g.
Optical Gravitational Lensing Experiment, OGLE), revolutionized our
knowledge about these important variables. They can no longer be
regarded as simple, radial-mode pulsators. New classes of multi-mode
RR~Lyr stars, that pulsate simultaneously in radial and non-radial
modes, were identified. In particular in RRc and RRd stars,
additional periodicities that fall in the $P_{\rm
x}/P_1\!\in\!(0.60,\,0.65)$ range were detected \citep[e.g.][and
references therein]{aqleo,om09,netzel3}. This form of pulsation must
be common among RRc/RRd stars as nearly all stars observed with the
utmost precision from space do show it \citep[14 out of 15 stars
observed, e.g.][]{szabo_corot,pamsm15,molnar,kurtz}. As period
ratios around 0.61 are the most common, following \cite{jurcsikM3b}
we will denote the group as \RRSO and the detected additional
frequencies as $\fSO$. In the recently proposed model, \cite{wd16}
assigns the additional variability with the harmonics of non-radial
modes of angular degrees 8 and 9. Non-radial modes themselves, were
also detected in several of these stars \citep[e.g.][initially
interpreted as period-doubling of the additional, $\fSO$
frequencies]{netzel3,pamsm15,molnar}. Another, even more puzzling
group, was revealed in the OGLE photometry of RRc stars. Here,
additional variability is of period longer than the first overtone
period; too long to be associated with the radial fundamental mode.
The period ratios tightly cluster around $P_1/P_{\rm x}\approx
0.686$ \citep{netzel2,ns16}. Peculiar examples of RRd stars were
also discovered \citep{rs15a,rs16,jurcsikM3a,anRRd_MC}; in
particular, \cite{anRRd_MC} identified a new class of anomalous RRd
stars, characterized by anomalous period ratios, in most cases lower
than in the ``classical'' RRd stars, by usual domination of the
fundamental mode and by common presence of the modulation.

Finally, the space photometry collected by {\it Kepler} and {\it
CoRoT} revolutionized our view on the Blazhko effect, a long-term
quasi-periodic modulation of pulsation amplitude and/or phase
\citep[for recent reviews see e.g.][]{geza,sm16bl}. Although the
effect is known for more than a century now, its origin remains a
mystery. Detection of period doubling in a significant fraction of
modulated RRab stars \citep[][]{kol10,szabo10,benko14} triggered the
development of new models explaining the phenomenon
\citep[e.g.][]{bk11,bryant16}. The nearly continuous and top-quality
photometry enabled a detailed study of the modulation on the
cycle-to-cycle basis \citep[e.g.][]{gugg,leborgne}.

The origin of the Blazhko effect or mechanisms behind the excitation
of various new forms of the multiperiodic RR~Lyr pulsation, remain
unclear. To test the existing models (and to propose the new ones),
new, top-quality observations are still needed. In particular, it is
important to determine how the incidence rates and properties of the
various pulsation forms depend on metallicity and on population
membership of the stars. In this respect, detailed studies of
globular cluster variables are crucial, but are very scarce. The
only globular cluster in which dynamical properties of RR~Lyr stars
were studied in detail is M3 \citep{jurcsikM3a,jurcsikM3b}. It is an
Oosterhoff I (OoI) cluster with mean metallicity of ${\rm
[Fe/H]}=-1.57$ belonging to a young halo population \citep{catelan}.
In this paper we analyse the photometry gathered by the Cluster AgeS
Experiment (CASE) project \citep{jka05} for another OoI cluster with
a significantly higher mean metallicity (${\rm [Fe/H]}=-0.95$),
belonging to the old halo population \citep{catelan}, NGC~6362.

NGC~6362 is a nearby ($\mu_V = 14.68$ mag) globular cluster located
away  from the Galactic disk at $b = -17^\circ.6$ \citep[2010
edition]{har10}. These properties together with a low concentration
make it an attractive target for studies with ground based
telescopes. Comprehensive CCD observations with 2.5-m du Pont and
1-m Swope telescopes of the Las Campanas Observatory were  performed
by \citet{mazur99}, \citet{olech01} and \citet{jka14}. It was found
that NGC~6362 hosts 35 RR~Lyr stars, whose basic parameters were
estimated by \citet{olech01} (O01 in the following) based on Fourier
decomposition of light curves. The data analyzed by O01 were
collected with the Swope telescope during only one observing season,
between April 17 and August 22, 1999. \citet{jka14} had a much
longer time-basis at their disposal, but they were mainly interested
in eclipsing binaries, and left the pulsating stars untouched.

In this paper we revisit the RR~Lyr stars of NGC~6362. Much longer
time-base and more numerous observations than used by O01, allow us
to conduct a detailed study of the dynamical properties of RR~Lyr
stars. Presence and properties of the Blazhko effect and excitation
of additional pulsation modes, both radial and non-radial, are at
the centre of our interests.

\section{Observational data and their analysis}\label{sect:data}

\begin{figure*}
\includegraphics[width=2\columnwidth]{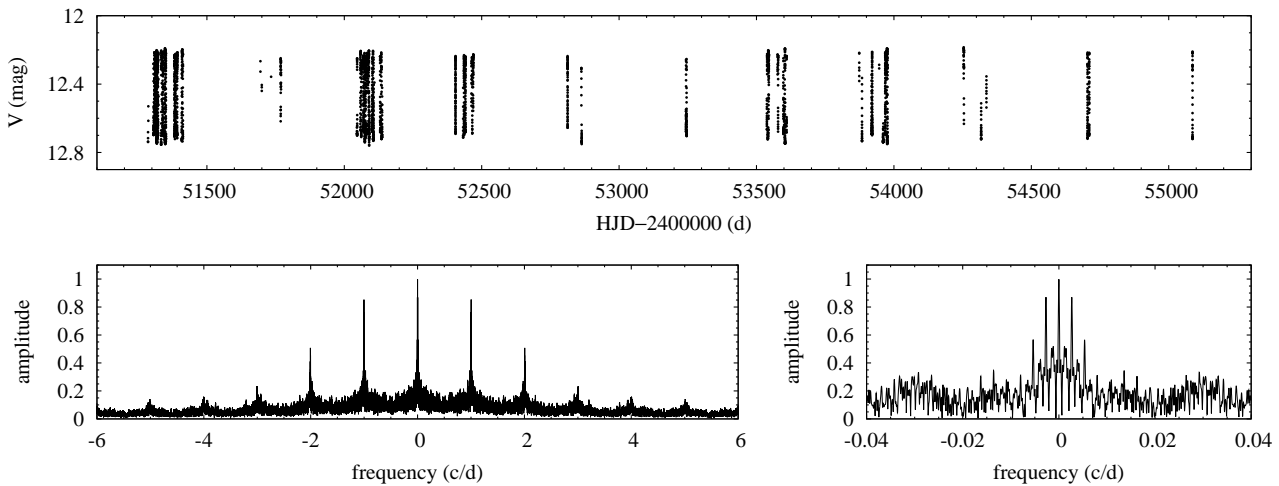}
\caption{A typical structure of the data gathered with the Swope
         telescope, illustrated with the example of variable V6 (top
         panel). The data clearly separate into 11 observing
         seasons. The bottom panel shows the spectral window (bottom
         left), including zoom into the fine structure of 1-yr
         aliases (bottom right).}
\label{fig:data}
\end{figure*}

Our paper is based on two sets of CCD images\footnote{ Data are
available at the website of the CASE project, http://case.camk.edu.pl}.
The first one was
obtained using the du Pont telescope and the 2048$\times$2048 TEK5
camera with a field of view of 8.84$\times$8.84 arcmin$^2$ and a
resolution of 0.259 arcsec/pixel. Observations were conducted on 45
nights from April 21, 1995 to September 26, 2009, always with the
same filters. Altogether, 1748 $V$-images were collected. The second
set was obtained with the Swope telescope and the 2048$\times$3150
SITE3 camera with a field of view of 14.8$\times$22.8 arcmin$^2$ and
a resolution of $0.435$\thinspace arcsec/pixel. Observations were
conducted on 103 nights from July 8, 1999 to September 9, 2009, also
keeping the same filters. For the analysis, 3200 $V$-images were
used. Since we are interested in the dynamical properties of RR~Lyr
stars, no transformation to the standard photometric system was
made; we use instrumental magnitudes through the paper. We refer the
interested reader to the study by O01 for the colour-magnitude
diagram of the cluster, mean magnitudes and colours of the discussed
RR~Lyr stars, and their physical properties derived from the $V$-
and $B$-band two-colour photometry through the relations given by
\cite{sicl} and \cite{geza98}.

Our analysis is based primarily on the Swope telescope data for two
reasons. First, for a given star they are typically twice as
numerous as the data collected with the du Pont telescope. Second,
data gathered with the du Pont telescope are of inferior quality as
compared to the data gathered with the Swope telescope.
Specifically, dispersion of the residuals of the Fourier fit to the
du Pont data is always higher, by at least 40 per cent. As a
consequence, when merging the Swope and the du Pont data the noise
level in the Fourier transform increases, making the detection of
weak, secondary periodicities more difficult. Lower quality of the
du Pont data might be due to different reduction procedures adopted
for the two telescopes: ISIS package of \cite{isis} was used for the
du Pont data and DIAPL package written by
W.~Pych\footnote{Procedures and the documentation are available at
http://users.camk.edu.pl/pych/DIAPL/index.html} was used for the
Swope data. Data from both telescopes were used together only to
study period changes of RR~Lyr variables. Inclusion of du Pont data
allowed us to add a few additional points to the phase-change (O--C)
diagrams.

Typical structure of data from the Swope telescope is illustrated in
the top panel of Fig.~\ref{fig:data}. The data were gathered over 11
observing seasons, below abbreviated as s1,\ldots, s11, clearly
visible in the figure. The number of observations vary from season
to season. The most extensive observations were gathered in s1
(typically $\sim\!1090$ data points), then in s3 ($\sim\!700$ data
points) and s8 ($\sim\!400$ data points). Data collected in other
seasons are less numerous (between $\sim\!40$ and $\sim\!300$ data
points). The data structure is nearly the same for all RR~Lyr stars
observed. The bottom panels of Fig.~\ref{fig:data} illustrate the
spectral window for data displayed in the top panel. Both daily and
1-yr aliases are very prominent and may become a source of confusion
during the analysis.

Our data analysis follows the standard consecutive prewhitening
technique. Significant periodicities are identified with the help of
the discrete Fourier transform and included in the sine series of
the following form:

\begin{equation}
v(t)=A_0+\sum_{i}A_i\sin\big(2\uppi \nu_i t +\phi_i\big)\,,\label{eq}
\end{equation}
which is fitted to the data with the help of non-linear least-square
method. Amplitudes, phases and frequencies are all adjusted
during the procedure. Prewhitened data are inspected for the
presence of additional, lower-amplitude periodicities, which are
iteratively included in eq.~\ref{eq}. We accept as significant
periodicities with signal-to-noise (S/N) above 4; the ratio is
determined from the frequency spectrum, with the noise evaluated as
a mean amplitude of the Fourier transform in the $0-10$\thinspace
c/d range. To account for the possible season-to-season zero-point
differences, ten independent offset coefficients are determined
during the procedure, by minimizing the dispersion of the fit (the
offsets are typically below $5$\thinspace mmag). Once the solution
converges (no new significant signals are detected), severe outliers
are removed from the data (5$\sigma$ clipping).

In eq.~\ref{eq} we include a few independent frequencies only. The
sine series describing a typical solution contains pulsation
frequency, $\nu$, and its harmonics, $k\nu$ (below, we will use
$\nu_0$ and $\nu_1$ for the fundamental mode and the first overtone
frequency, respectively; at the moment, there is no need to
differentiate the two). For the single-periodic and non-modulated
star, these are the only terms in eq.~\ref{eq}, which reduces to a
finite order Fourier series. Quite often, however, we detect
modulation of pulsation (the Blazhko effect) or multiperiodic
pulsation (or sometimes both). In the frequency spectrum, the
modulation manifests as equally spaced multiplets centred at the
frequency of the radial mode and its harmonics. Hence, only one
independent frequency, modulation frequency, $\nu_{\rm m}$, is added
to the solution. Components of the multiplet are then described as
$k\nu\pm l\nu_{\rm m}$. Modulation period is simply $P_{\rm
m}=1/\nu_{\rm m}$. Frequencies of additional periodicities, related
to e.g. additional radial/non-radial pulsation modes are denoted as
$\nu_{\rm x}$. Linear combinations with the dominant radial
pulsation frequency, typically of the form $k\nu\pm \nu_{\rm x}$,
are often detected and included in the solution.

Slow phase and/or amplitude changes may occur in RR~Lyr stars; in
fact they are rather common in the RRc variables. Phase changes are
usually most pronounced and often of irregular nature. As a result,
the peaks detected at the radial mode frequency (and harmonics) are
non-coherent and a residual unresolved power remains in the
frequency spectrum after prewhitening. The resulting, increased
noise level in the Fourier transform may hamper the detection of
additional, low-amplitude periodicities. To circumvent the problem,
two techniques were applied. In the first one, we simply focus on a
short part of the data, either corresponding to the first
season, s1, or to the first four seasons, s1-s4. On a shorter time
scale, the phase changes are not that disruptive. The second
technique we apply is a time-dependent prewhitening. The details of
the method and its application to quasi-continuous {\it Kepler} data
are discussed in \cite{pamsm15}, the Appendix. In \cite{netzel1} the
technique was applied to seasonal, ground-based data and here we
follow the same approach. In a nutshell, the data are first divided
into seasons, and amplitudes and phases are determined for each
season separately. Then, the data are prewhitened with the sine
series of the form displayed in eq.~\ref{eq}, except that amplitudes
and phases are now season-dependent. The procedure filters out
possible amplitude/phase variations on a time scale longer than the
typical length of the observing season. The frequency spectrum is
cleaned from the unresolved power and the overall noise level drops,
enabling detection of additional low-amplitude periodicities.
Because of scarce data in some of the observing seasons, the
technique can in most cases be applied to five seasons only (s1, s3,
s4, s7 and s8).

Whenever these techniques appeared crucial for the detection of
additional phenomena (modulation, additional periodicities) we
explicitly state so while discussing a given star.

\section{Results}\label{sect:results}

\subsection{Overview}

We have analysed photometric data for 35 pulsators from NGC~6362,
all of which were identified as RR~Lyr stars in the study of O01 and
earlier studies of the cluster summarized there. Of these, 18 were
classified as RRab and 17 as RRc variables. For the majority of
stars we confirm the classification of O01, with the exception of
V3, V34 and V37. In the first two stars, previously classified as
RRab, we detect the first overtone, and now classify them as RRd
variables; in fact they are new members of the anomalous RRd class
identified recently by \cite{anRRd_MC}, as we discuss in detail in
Section~\ref{ssect:anRRd}. In addition, V37, classified by O01 as
RRc, is discussed separately in our study (Section~\ref{ssect:V37}).
Most likely, V37 is not RR~Lyr star, but a pulsating variable of
different nature. The basic properties of 16 RRab and 16 RRc
variables are collected in Tab.~\ref{tab:abc} (which also contains
data on V37 in the last row). These include pulsation period and
pulsation amplitude as well as information about additional
phenomena detected in the stars (the last column). The phenomena
are: Blazhko modulation (`BL' in the last column), double-periodic
Blazhko modulation (`2$\times$BL') and the presence of the
additional, non-radial pulsation (`nr'). All 11 seasons of Swope
data were used to derive pulsation periods and amplitudes given in
Tab.~\ref{tab:abc} and hence, these quantities are means over the
time span of the observations. This comment is particularly relevant
for the RRc stars, where we often detect fast and irregular period
changes (see Section~\ref{ssect:rrc}).

\begin{table}
\centering
\caption{Basic properties of the analysed RRab and RRc variables: star's
         id, type, pulsation period and Fourier amplitude, i.e.
         $A_i$ from eq. 1 at the fundamental/first overtone
         frequency. All numbers are given to the last significant
         digit. In the last column remarks are given: `BL' --
         Blazhko effect detected, `2$\times$BL' -- double-periodic
         Blazhko effect detected, `nr' -- non-radial mode(s)
         detected. Properties of the two RRd stars are listed
         separately in Tab.~\ref{tab:d}.}

\label{tab:abc}
\begin{tabular}{lllll}
id  & type & $P$ (d)     & $A$ (mag) & remarks \\
\hline
V1  & RRab & 0.50479162  & 0.3667 & BL  \\
V2  & RRab & 0.488973010 & 0.4143 &     \\
V5  & RRab & 0.52083783  & 0.3039 & BL  \\
V7  & RRab & 0.521581388 & 0.3592 & BL  \\
V12 & RRab & 0.5328814   & 0.3285 & BL  \\
V13 & RRab & 0.58002740  & 0.3556 & BL  \\
V16 & RRab & 0.525674215 & 0.3458 &     \\
V18 & RRab & 0.51288484  & 0.3429 & BL  \\
V19 & RRab & 0.59450528  & 0.2080 &     \\
V20 & RRab & 0.69835898  & 0.1535 & BL  \\
V25 & RRab & 0.455890887 & 0.4195 &     \\
V26 & RRab & 0.60217449  & 0.2132 &     \\
V29 & RRab & 0.64778329  & 0.1687 & BL  \\
V30 & RRab & 0.61340457  & 0.3241 & 2$\times$BL \\
V31 & RRab & 0.60021294  & 0.2211 & BL  \\
V32 & RRab & 0.49724171  & 0.3759 & BL  \\
\hline
V6  & RRc  & 0.26270671  & 0.2226 & 2$\times$BL \\
V8  & RRc  & 0.38148471  & 0.2023 & nr  \\
V10 & RRc  & 0.265638816 & 0.2222 & BL  \\
V11 & RRc  & 0.288789268 & 0.2433 &     \\
V14 & RRc  & 0.24620647  & 0.1587 &     \\
V15 & RRc  & 0.279945707 & 0.2194 & nr  \\
V17 & RRc  & 0.31460473  & 0.1995 & nr  \\
V21 & RRc  & 0.281390043 & 0.2366 & nr  \\
V22 & RRc  & 0.26683523  & 0.2304 &     \\
V23 & RRc  & 0.275105063 & 0.2428 & nr  \\
V24 & RRc  & 0.32936190  & 0.2218 & nr  \\
V27 & RRc  & 0.27812399  & 0.2402 & nr  \\
V28 & RRc  & 0.3584133   & 0.1898 &     \\
V33 & RRc  & 0.30641758  & 0.2007 & nr  \\
V35 & RRc  & 0.29079074  & 0.2098 & nr  \\
V36 & RRc  & 0.31009148  & 0.1790 & nr, BL \\ 
\hline
V37 & ?    & 0.25503903  & 0.1174 & BL  \\
\hline
\end{tabular}
\end{table}

\begin{figure}
\includegraphics[width=\columnwidth]{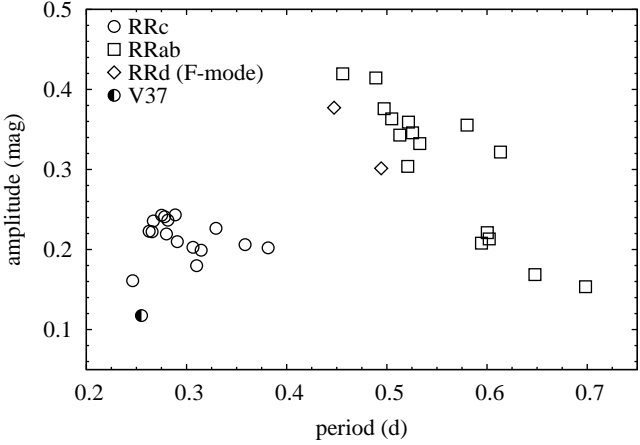}
\caption{Period-amplitude diagram for RR~Lyr stars of NGC~6362. Note
         that Fourier amplitudes are used in the plot.}
\label{fig:bailey}
\end{figure}

\begin{figure}
\includegraphics[width=\columnwidth]{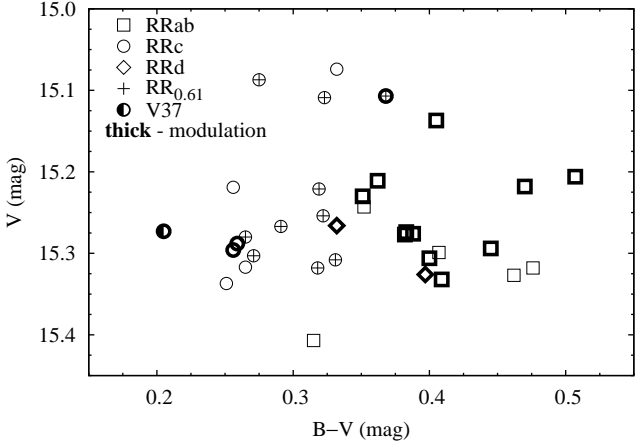}
\caption{Colour-magnitude diagram for RR~Lyr stars of NGC~6362.
         Standard magnitudes and colours were adopted from O01.}
\label{fig:cmd}
\end{figure}

\begin{figure}
\includegraphics[width=\columnwidth]{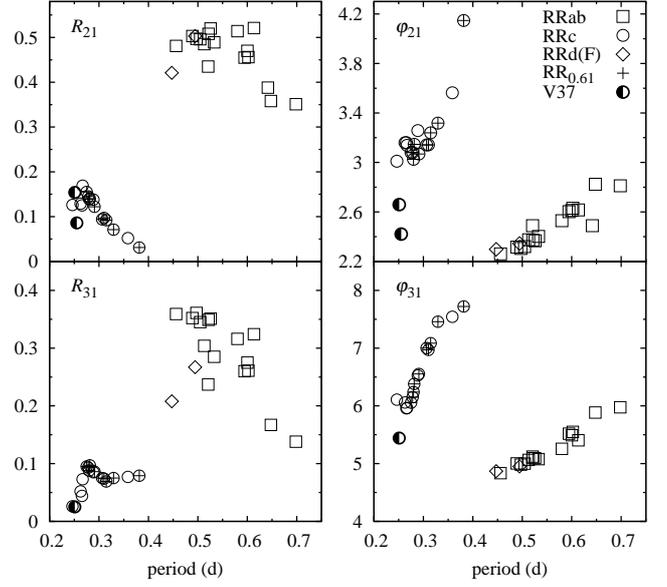}
\caption{Low-order Fourier decomposition parameters for pulsating
         stars of NGC~6362.}
\label{fig:fp}
\end{figure}

\begin{figure*}
\includegraphics[width=2\columnwidth]{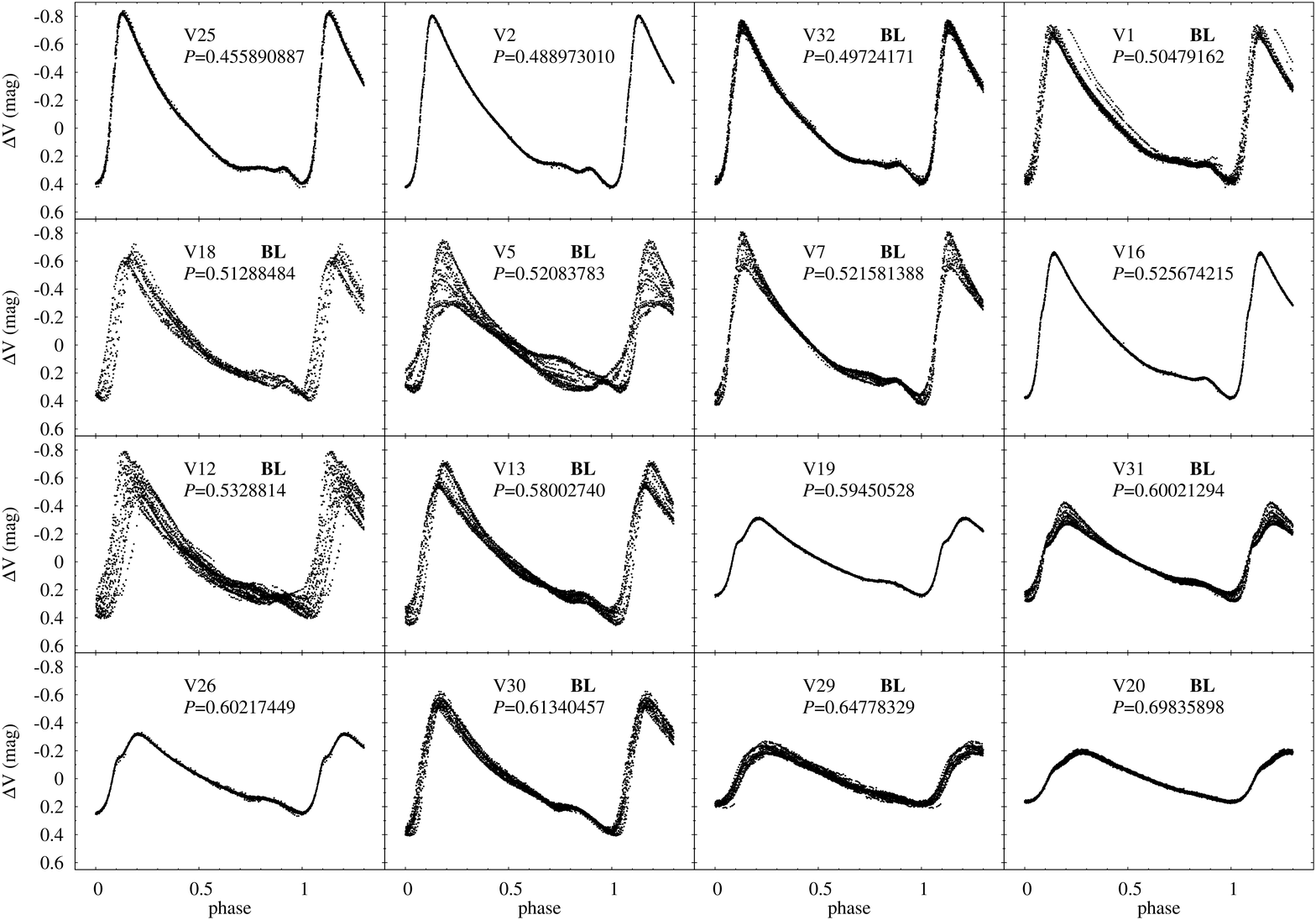}
\caption{A collection of phased light curves for all observed RRab
         stars, sorted by the increasing pulsation period. Data from
         all 11 observing seasons were used for each star. The
         adopted pulsation period is given in each panel, together
         with information about possible modulation (BL).}
\label{fig:lc_rrab}
\end{figure*}

\begin{figure*}
\includegraphics[width=2\columnwidth]{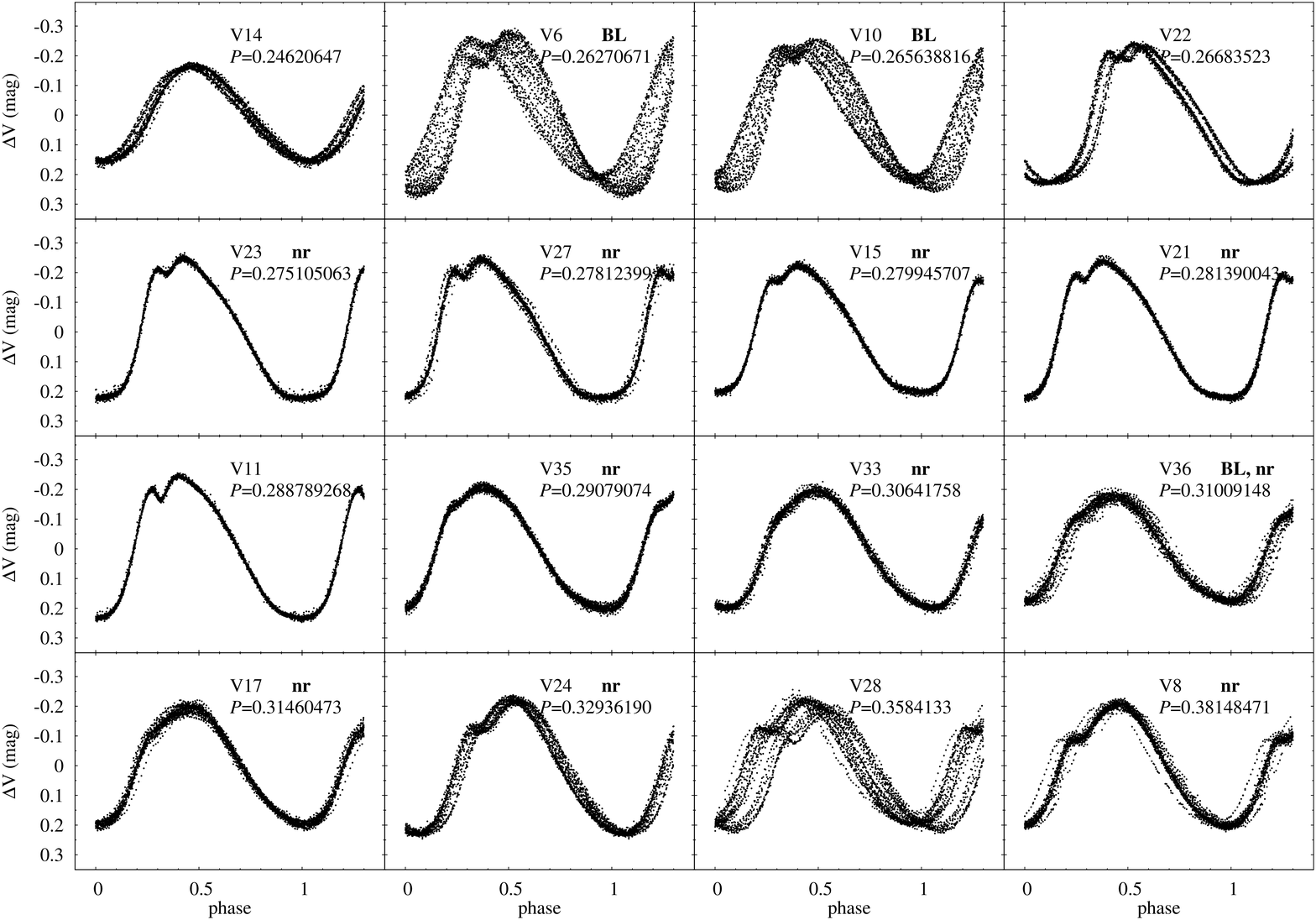}
\caption{A collection of phased light curves for all observed RRc
         stars, sorted by the increasing pulsation period. Data from
         all 11 observing seasons were used for each star. The
         adopted pulsation period is given in each panel, together
         with information about possible modulation (BL) or about
         presence of additional non-radial pulsation (nr).}
\label{fig:lc_rrc}
\end{figure*}

In Fig.~\ref{fig:bailey} we present the period-amplitude diagram for
the discussed stars, which nicely confirms the adopted mode
classification. For the two RRd stars, data corresponding to the
dominant fundamental mode are used in this plot. Colour-magnitude
diagram is presented in Fig.~\ref{fig:cmd}. For this plot we used
the standard $V$ and $B$-band mean brightness from O01 (their
tab.~1). A separation of RRc from RRab variables is rather clear. Of
the two newly identified RRd stars, V3 is located in the
colour-magnitude plot in between RRc and RRab stars. V34 appears a
bit cooler, but its location does not arise suspicions. Further
support for the adopted mode classification is presented in
Fig.~\ref{fig:fp}, which presents low-order Fourier decomposition
parameters for all the discussed variables. RRab and RRc stars
clearly separate in these plots. The two RRd stars, with dominant
pulsation in the fundamental mode, fit within the RRab group (no
harmonics are detected for the first overtone, consequently, the
corresponding Fourier parameters cannot be computed). In
Figs.~\ref{fig:lc_rrab} and \ref{fig:lc_rrc} we present a gallery of
phased light curves of RRab and RRc variables, respectively. Stars
are sorted by the increasing pulsation period (given in each panel).
Note that for all stars, data from all 11 seasons were used in the
plots. For some stars the Blazhko modulation is obvious (e.g.
in V5 and V31, which are of RRab type, or in V6 and V10, which are
of RRc type). In others, particularly in several RRc stars, the data
do not phase well, but form a band of light curves shifted in phase
(e.g. V8, V14, V24, V28). The bands are due to period changes in
these stars. The most extreme case is V28.

In the following Sections, we first discuss RRab variables
(Section~\ref{ssect:ab} and \ref{ssec:notesab}), then RRc stars
(Section~\ref{ssect:rrc} and \ref{ssect:notesRRc}), two RRd
pulsators (Section~\ref{ssect:rrd}) and finally the oddball, V37
(Section~\ref{ssect:V37}).


\subsection{RRab stars}\label{ssect:ab}

Altogether, NGC~6362 hosts 16 RRab stars; their basic properties are
given in Tab.~\ref{tab:abc}, and phased light curves are presented
in Fig.~\ref{fig:lc_rrab}.

\begin{figure*}
\includegraphics[width=2\columnwidth]{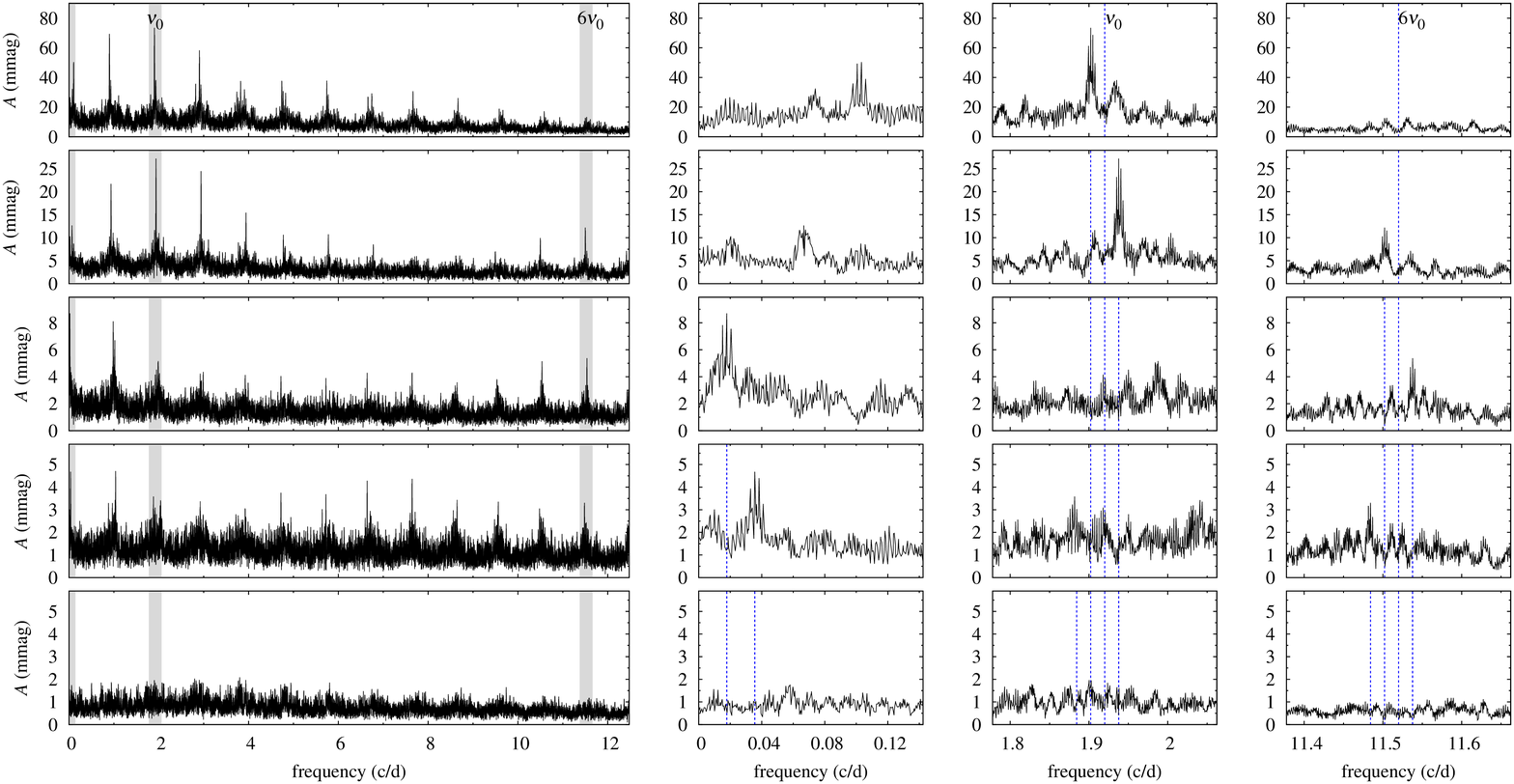}
\caption{Prewhitening sequence for Blazhko RRab variable V5. The
         first column shows the wide frequency range, up to
         $6.5\nu_0$. The next columns show zooms into three
         grey-shaded regions: at a low frequency range, at a range
         centred at $\nu_0$, and at a range centred at $6\nu_0$.
         Consecutive rows show, from top to bottom: ({\it first
         row}) frequency spectrum after prewhitening with the
         fundamental mode and its harmonics; ({\it second}) in
         addition $k\nu_0-\nu_{\rm m}$ for $k=1,...,5$ were
         prewhitened; ({\it third}) in addition $k\nu_0-\nu_{\rm m}$
         for $k=6,...,10$ and $k\nu_0+\nu_{\rm m}$ for $k=1,...,5$
         were prewhitened; ({\it fourth}) all significant triplet
         components and the significant peak at $\nu_{\rm m}$ were
         prewhitened; ({\it fifth}) all significant quintuplet
         components and the significant peak at $2\nu_{\rm m}$ were
         prewhitened. Except the first-column plots, prewhitened
         frequencies are marked with dashed lines.}
\label{fig:05bl}
\end{figure*}

Eleven variables show the Blazhko modulation, which was recognized
based on the analysis of the frequency spectrum. For some stars
however, the effect is obvious already from the analysis of the
phased light curve (see e.g. V5, V7, V13 or V31 in
Fig.~\ref{fig:lc_rrab}). The prewhitening sequence is illustrated
with the example of variable V5 which shows clear, large-amplitude
modulation of pulsation with a period of about $56.2$\thinspace d.
Fig.~\ref{fig:05bl} shows a large section of the frequency spectrum
of V5 covering the first five harmonics (left-most column), and
zooms into smaller sections: around modulation frequency (second
column), centred at $\nu_0$ (third column) and centred at $6\nu_0$
(last column). The top row in Fig.~\ref{fig:05bl} illustrates the
frequency spectrum after prewhitening with the fundamental mode
frequency and its harmonics. A triplet at $\nu_0$ is apparent, with
a higher amplitude side peak on the low-frequency side of $\nu_0$.
The following rows of Fig.~\ref{fig:05bl} show the prewhitening
process, first with triplet components (second, third and fourth
rows), and finally with all significant quintuplet components, which
are more pronounced at higher order harmonics of the fundamental
mode. Quintuplets are incomplete; only negative frequency
components, i.e., $k\nu_0-2\nu_{\rm m}$, are significant. We also
note that in the low frequency range, peaks at the modulation
frequency and twice the modulation frequency are prominent (second
column in Fig.~\ref{fig:05bl}). After prewhitening with the
fundamental mode frequency, its harmonics, and all multiplet
components no additional significant power is detected in the
frequency spectrum (bottom row of Fig.~\ref{fig:05bl}).

The top section of Tab.~\ref{tab:bl} summarizes the basic properties
of the Blazhko modulation in RRab stars of NGC~6362. These are:
modulation period, $P_{\rm m}$, amplitudes of the lower ($A_-$) and
higher ($A_+$) frequency side peaks at $\nu_0$, relative amplitude
of modulation, i.e. $A_{\rm mod}\!=\!{\rm max}(A_-,\,A_+)/A$, the
asymmetry parameter defined by \cite{alcock} as,
$Q=(A_+-A_-)/(A_++A_-)$, indication which data were used in the
analysis (`data' column) and some details on the main modulation
components detected in the spectrum.

In all but one modulated RRab stars only a single modulation period
was found. It varies from $\approx\!17$\thinspace d (V1, V32) to
$\approx\!82$\thinspace days (V31). Two modulation periods were
identified in V30, $\approx\!34.8$\thinspace d and
$\approx\!216.4$\thinspace d. The relative modulation amplitude vary
from $\approx\!3$\thinspace per cent (V30, V32) to
$\approx\!24$\thinspace per cent (V5). Clear modulation triplets
were detected in the majority of stars, with the exception of V12
and V20, in which only close doublets were detected at $k\nu_0$.
Complete quintuplets or components of the quintuplets were detected
in V1, V5, V13 and V29. V29 is an interesting case. No additional
side peaks were detected on the low frequency side of $k\nu_0$,
however side peaks with $+\nu_{\rm m}$ and $+2\nu_{\rm m}$
separation appear on the high frequency side of $k\nu_0$. In
addition, starting from $2\nu_0$, the $+2\nu_{\rm m}$ components are
significantly higher than the $+\nu_{\rm m}$ components. Thus,
incomplete and strongly asymmetric quintuplets are detected in V29.
Finally, in V13 we detect one significant component of septuplet,
$10\nu_0-3\nu_{\rm m}$. In four stars, in the low-frequency range we
detect significant peaks at the modulation frequency (V5, V7, V13,
V31) and at its harmonic (V5, V13).

V30 is an interesting and rare case of RRab star with two modulation
periods. The modulation with the shorter period is dominant -- the
relative modulation amplitude is 7.5 per cent, to be compared with
2.5 per cent for the modulation with the secondary, longer period.
For both modulation periods, clear triplet structures are detected
at $k\nu_0$.

We note that O01 did not analysed the Blazhko effect in RRab stars
of NGC~6362 in any detail. Based on the light curve scatter, V5, V7,
V12, V13 and possibly V18, were regarded as stars showing the
Blazhko effect. A further discussion on the Blazhko effect in
NGC~6362 is postponed to Section~\ref{ssect:blazhko}.

With the exception of V25 (see Section~\ref{ssec:notesab} below) no
additional periodicities that could be associated with other radial
or non-radial modes of pulsation were detected in RRab stars.

\begin{table*}
\centering
\caption{Properties of the Blazhko variables. Consecutive
         columns contain: star's id, type, pulsation period,
         modulation period, amplitudes of the modulation side peaks
         located near the basic pulsation frequency, at lower
         ($A_{-}$) and at higher frequency ($A_{+}$), relative
         modulation amplitude, $A_{\rm mod}$, and asymmetry
         parameter, $Q$. Results are based on the analysis of data
         indicated in the 9th column. The last column lists the main
         modulation side peaks detected in the frequency spectrum.}
\label{tab:bl}
\begin{tabular}{llllrrrrlp{5.8cm}}
id      & type & $P$         & \,~$P_{\rm m}$ & $A_{-}$  & $A_{+}$  & $A_{\rm mod}$ & $Q$ & data & main modulation components\\
        &      & (d)         & \,~(d)         & (mag)    & (mag)    &               &     &      & \\
\hline
V1      & RRab & 0.50479165  &   \,~17.4043  & $0.0065$ & $0.0182$ & $0.050$ & $ 0.477$ &  s1-s4 & $\pm\nu_{\rm m}$ components at $\nu_0,\ldots,7\nu_0$; $+\nu_{\rm m}$ components at $8\nu_0,\ldots,10\nu_0$; $+2\nu_{\rm m}$ components at $3\nu_0,5\nu_0$ and weak at $6\nu_0$\\ 
V5      & RRab & 0.52083783  &   \,~56.190   & $0.0723$ & $0.0307$ & $0.238$ & $-0.404$ &  all   & $\nu_{\rm m}$; $2\nu_{\rm m}$; $\pm\nu_{\rm m}$ components at $\nu_0,\ldots,7\nu_0$, $9\nu_0$, $10\nu_0$; $-\nu_{\rm m}$ components at $8\nu_0$, $11\nu_0$; $-2\nu_{\rm m}$ components at $3\nu_0,\ldots, 10\nu_0$\\  
V7      & RRab & 0.521581388 &   \,~41.191   & $0.0239$ & $0.0143$ & $0.067$ & $-0.253$ &  all   & $\nu_{\rm m}$; $\pm\nu_{\rm m}$ components at $\nu_0,\ldots,13\nu_0$;  $+\nu_{\rm m}$ components at $14\nu_0,\ldots,19\nu_0,21\nu_0,22\nu_0$ \\  
V12     & RRab & 0.53286906  &   \,~57.710   & $0.0548$ &  -       & $0.164$ & $-1.000$ &  s1-s4 & $-\nu_{\rm m}$ components at $\nu_0,\ldots,8\nu_0$\\ 
V13     & RRab & 0.58002740  &   \,~35.2368  & $0.0344$ & $0.0320$ & $0.097$ & $-0.036$ &  all   & $\nu_{\rm m}$; $2\nu_{\rm m}$; $\pm\nu_{\rm m}$ components at $\nu_0,\ldots,9\nu_0$; $-\nu_{\rm m}$ component at $10\nu_0$; $\pm2\nu_{\rm m}$ components at $4\nu_0,\ldots,8\nu_0$; $-2\nu_{\rm m}$ components at $9\nu_0,\ldots,13\nu_0$; $+2\nu_{\rm m}$ components at $2\nu_0, 3\nu_0$; $-3\nu_{\rm m}$ component at $9\nu_0,\ldots,13\nu_0$ \\
V18     & RRab & 0.51288484  &   \,~78.036   & $0.0286$ & $0.0486$ & $0.142$ & $ 0.259$ &  all   & $\pm\nu_{\rm m}$ components at $\nu_0,\ldots,6\nu_0$; $+\nu_{\rm m}$ components at $7\nu_0$, $8\nu_0$\\ 
V20     & RRab & 0.69835898  &   \,~73.332   &  -       & $0.0089$ & $0.058$ & $ 1.000$ &  all   & $+\nu_{\rm m}$ components at $\nu_0,\ldots,5\nu_0$\\
V29     & RRab & 0.64778329  &   \,~66.917   &  -       & $0.0209$ & $0.124$ & $ 1.000$ &  all   & $+\nu_{\rm m}$ components at $\nu_0,\ldots, 3\nu_0$; $+2\nu_{\rm m}$ components at $2\nu_0,\ldots, 4\nu_0$; \\
V30     & RRab & 0.61340457  &   \,~34.809   & $0.0241$ & $0.0109$ & $0.074$ & $-0.376$ &  all   & $\pm\nu_{\rm m1}$ components at $\nu_0,\ldots,6\nu_0$; $-\nu_{\rm m1}$ component at $7\nu_0$\\
        &      &             &     216.37    & $0.0066$ & $0.0081$ & $0.025$ & $ 0.101$ &        & $\pm\nu_{\rm m2}$ components at $\nu_0,\ldots,5\nu_0$; $+\nu_{\rm m2}$ component at $6\nu_0$\\
V31     & RRab & 0.60021294  &   \,~82.267   & $0.0191$ & $0.0133$ & $0.086$ & $-0.179$ &  all   & $\nu_{\rm m}$; $\pm\nu_{\rm m}$ components at $\nu_0,\ldots,16\nu_0$, $18\nu_0,\ldots,20\nu_0$; $+\nu_{\rm m}$ components at $17\nu_0$, $21\nu_0$\\
V32     & RRab & 0.49724171  &   \,~17.3010  & $0.0053$ & $0.0119$ & $0.032$ & $ 0.384$ &  all   & $\pm\nu_{\rm m}$ at $\nu_0$; $+\nu_{\rm m}$ components at $2\nu_0,\ldots,12\nu_0$\\ 
\hline
V6      & RRc  & 0.26270529  &   \,~15.4504  & $0.0610$ & $0.0624$ & $0.280$ &  $0.012$ &  s1-s4 & $2\nu_{\rm m1}$; $\pm\nu_{\rm m1}$ components at $\nu_1,\ldots,6\nu_1$; $+\nu_{\rm m1}$ component at $8\nu_1$; $\pm2\nu_{\rm m1}$ components at $2\nu_1,4\nu_1,\ldots,7\nu_1$; $-2\nu_{\rm m1}$ component at $3\nu_1$\\
        &      &             &   \,~13.771   &  -       & $0.0033$ & $0.015$ &  $1.000$ &        & $+\nu_{\rm m2}$ at $\nu_1$\\
V10     & RRc  & 0.265638816 & \,~\,~8.52242 & $0.0429$ & $0.0544$ & $0.245$ &  $0.118$ &  all   & $\nu_{\rm m}$; $2\nu_{\rm m}$; $\pm\nu_{\rm m}$ components at $\nu_1,\ldots,7\nu_1$; $+\nu_{\rm m}$ component at $8\nu_1$;   $\pm2\nu_{\rm m}$ component at $2\nu_1$; $-2\nu_{\rm m}$ components at $3\nu_1,\ldots,5\nu_1$ \\
V36     & RRc  & 0.310084    &   \,~13.764   &  -       & $0.0036$ & $0.020$ &  $1.000$ &  s1    & $+\nu_{\rm m}$ at $\nu_1$ \\
\hline
\end{tabular}
\end{table*}

\subsection{Notes on individual RRab stars}\label{ssec:notesab}

\noindent{\it V16} -- in the analysis of all data, after
prewhitening with the fundamental mode frequency and its harmonics,
we detect significant (${\rm S/N}\approx 5$) remnant power at
$k\nu_0$ in the form of several closely spaced peaks of similar
height, which may be due to a long-period modulation. The period
would be either $\approx\!1070$\thinspace d or
$\approx\!576$\thinspace d, depending which of the 1-yr aliases is
chosen. Although these are formally resolved, the frequency spectrum
at $k\nu_0$ corresponds to a wide power excess rather than to a
genuine modulation and may result from irregular amplitude and or
phase changes. In addition, after applying the time-dependent
prewhitening to the data of the most numerous seasons (s1, s3, s4,
s7 and s8) no additional power is detected at the fundamental mode
and its harmonics. Consequently, we do not regard V16 as Blazhko
variable.

\smallskip

\noindent{\it V25} -- in the analysis of all data, the fundamental
mode and its harmonics are non-coherent -- a strong remnant power
remains at $k\nu_0$ after prewhitening. Interestingly, when we
analyse the first four seasons only, the effect disappears: the
fundamental mode and its harmonics are coherent and no remnant power
is present. This suggest a possible significant period change/jump
during the observations. In the residuals of the s1--s4 data we
detect an additional periodicity (${\rm S/N}=6.1$) of long period,
$P_{\rm x}=1.03503$\thinspace d, $P_{\rm x}/P_0=2.2704$. This
periodicity cannot correspond to any acoustic mode of oscillation.
No combination frequencies with the radial mode are detected.
In addition, we note that a signal of nearly the same frequency is
detected also in two other stars, in V10 and in V11 (see
Section~\ref{ssect:notesRRc}). Consequently, this signal cannot be
intrinsic to the stars, but is most likely an artefact (origin of
which remains unknown).

\smallskip

\noindent{\it V2, V19 and V26} -- these stars are genuine,
single-periodic RRab variables, without any sign of modulation or
additional periodicities in the frequency spectrum.

\subsection{RRc stars}\label{ssect:rrc}

Altogether, NGC~6362 hosts 16 genuine RRc stars; their basic
properties are given in Tab.~\ref{tab:abc}, and phased light curves
are presented in Fig.~\ref{fig:lc_rrc}. V37, previously classified
as RRc, is discussed separately in Section~\ref{ssect:V37}.

The most interesting result is the detection of additional
periodicities in the $P/P_1\!\in\!(0.60,\,0.65)$ range in 10 out of
16 RRc stars. These are new members of the \RRSO class, described in
the Introduction. Data on all \RRSO stars are collected in
Tab.~\ref{tab:61}, in which we report: period of the additional
variability, $P_{\rm x}$, period ratio with the first overtone
period, and amplitude of the additional variability, $A_{\rm x}$.
This amplitude is always low, in the mmag range (the highest
amplitude is $\approx\!5$ mmag in V17), which makes the detection of
additional variability challenging.  This is particularly true for
stars in which strong phase changes are present. In such a case,
after prewhitening with the first overtone frequency and its
harmonics, residual power remains in the frequency spectrum and the
overall noise level in the Fourier transform is increased. This
hampers the detection of the possible secondary signal. For such
stars we can circumvent the problem, either by selecting a shorter
data subset, so the period change is not as pronounced, or by
applying time-dependent prewhitening. In the former case, we select
either the first season (s1), or the first four seasons (s1-s4).
That way we include the most densely sampled seasons s1, s3 and s4
(see Fig.~\ref{fig:data}) which guarantees the lowest noise level in
the Fourier transform. Sixth column of Tab.~\ref{tab:61} explicitly
identifies which data were used in the analysis. Time-dependent
prewhitening was applied for three stars (`tdp' in the last column
of Tab.~\ref{tab:61}).

\begin{figure*}
\includegraphics[width=2\columnwidth]{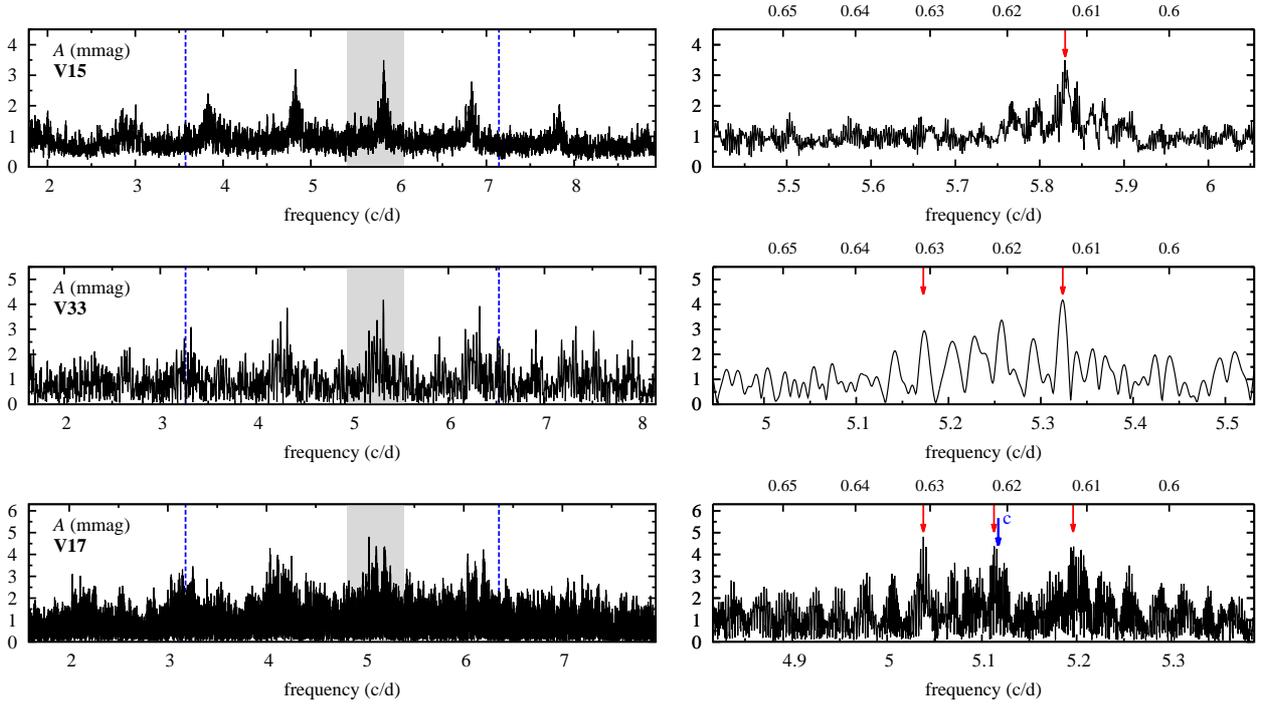}
\caption{Detection of additional periodicities in three RRc stars,
         V15, V33 and V17. Left panels show the frequency spectrum
         in the $(0.5\fO,\,2.5\fO)$ range, prewhitened with the
         first overtone frequency and its harmonics (dashed lines).
         Right panels show zoom into the frequency range with
         additional peaks, marked with grey-shaded box in the left
         panels. Detected peaks are marked with arrows. At the top
         axis of right panels, period ratio scale, $P/P_1$ is
         plotted.}
\label{fig:61freq}
\end{figure*}

The detections are illustrated in Fig.~\ref{fig:61freq} for three
stars, V15, V33 and V17, in the top, middle and bottom rows,
respectively. The left panels show large sections of the frequency
spectrum after prewhitening with the first overtone frequency and
its harmonics. Additional, significant signals are clearly detected
between $\fO$ and $2\fO$. The right panels of Fig.~\ref{fig:61freq}
show zooms into the interesting part of the spectrum. For V15 a
power excess of a complex structure is well visible. After
prewhitening with the frequency of the highest peak, located at
$P_{\rm x}/P_1\!\approx\!0.613$, close significant peaks still
remain in the spectrum. For V33 (middle row of
Fig.~\ref{fig:61freq}), after prewhitening with the highest peak
located at $P_{\rm x}/P_1\!\approx\!0.613$ (marked with an arrow),
another, well separated peak becomes significant (at $P_{\rm
x}/P_1\!\approx\!0.631$, also marked with an arrow). In the case of
V17 (bottom row of Fig.~\ref{fig:61freq}), three well separated
peaks are detected (centred at $P_{\rm
x}/P_1\!\approx\!0.612,\,0.622$ and $0.631$). V17 and V33 are the
only stars in which more than one peak is reported in
Tab.~\ref{tab:61}. These peaks correspond to two (V33) or three
(V17) sequences formed by \RRSO stars in the Petersen diagram (see
the discussion in Section~\ref{ssect:nr}). For other stars, only one
entry is present in Tab.~\ref{tab:61}. This doesn't mean that the
corresponding peak is coherent. Typically, the detected peak has a
complex structure, similar to that illustrated for V15 in the top
panel of Fig.~\ref{fig:61freq}. In all such cases, only the
frequency and amplitude of the highest peak is reported in
Tab.~\ref{tab:61}. The described, complex appearance of additional
periodicities in the frequency spectrum is common for this class of
stars, as we discuss in more detail in Section~\ref{ssect:nr}.

Another interesting phenomenon, detected in three RRc stars, V6, V10
and V36, is the Blazhko effect. Basic properties of the modulation
are given in the bottom section of Tab.~\ref{tab:bl}. Modulation
periods are significantly shorter than those observed in RRab stars;
the longest is only $\approx\!15.5$\thinspace days (V6), while the
shortest modulation period is $\approx\!8.5$\thinspace d (V10). In
all three stars the modulation side peaks on higher frequency side
of $\nu_1$ are higher; although for V6 and V10 the two detected
triplet components are of comparable height.

There is no doubt that a genuine modulation is observed in V6 (with
$P_{\rm m1}$) and in V10. In both cases at least one complete
quintuplet is observed at the harmonics of the first overtone
frequency. Admittedly, the secondary modulation in V6 and modulation
in V36 are less firmly established. Although in both cases we detect
a significant side peak on the higher frequency side of the first
overtone frequency, it is the only additional signal detected in the
frequency spectrum, i.e. no additional modulation side peaks are
identified.\footnote{We also note that the two modulation periods
are the same within the resolution of the data. The same modulation
periods and similar appearance of the side peaks in the frequency
spectrum (on the positive side of $\nu_1$) may suggest an artificial
origin of this modulation in the two stars; however it is hard to
postulate any reasonable cause of such an artefact.}

We note that V36 shows the Blazhko effect and $\fSO$ periodicity,
simultaneously. Further discussion of the Blazhko effect and of
\RRSO stars in NGC 6363 is in Section~\ref{ssect:blazhko} and
\ref{ssect:nr}, respectively.

\begin{figure*}
\includegraphics[width=2\columnwidth]{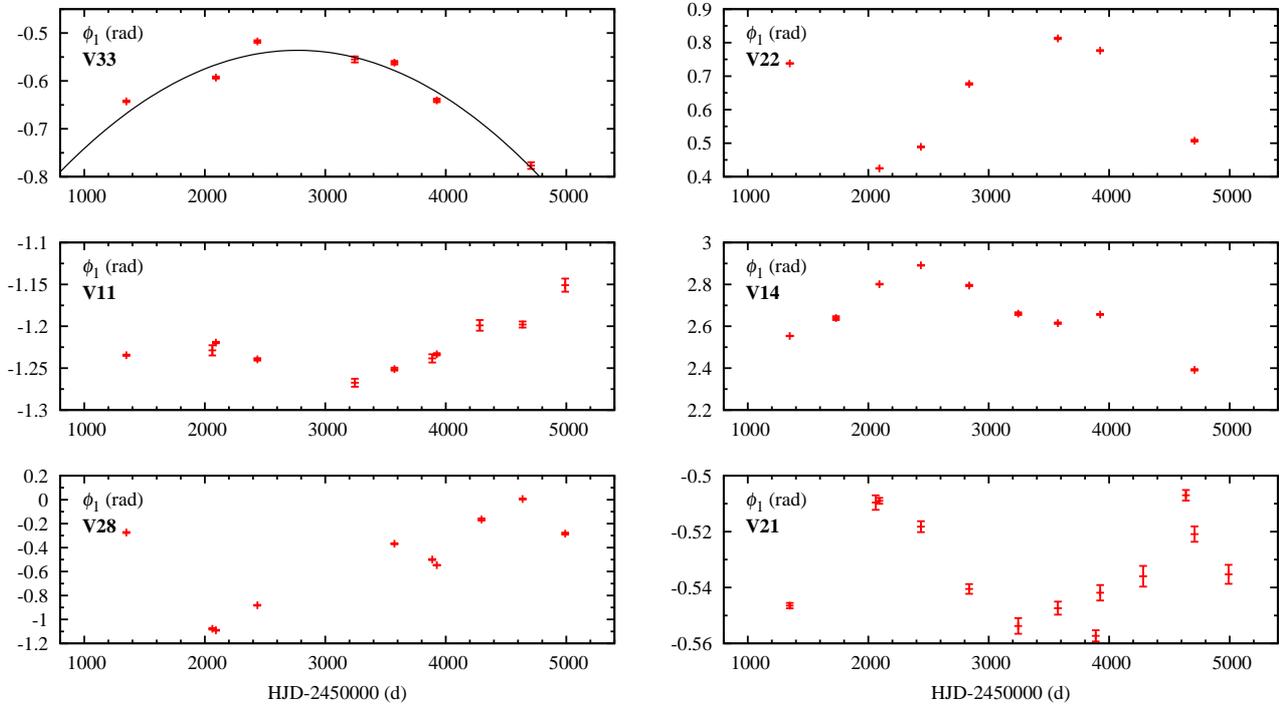}
\caption{Phase changes in 6 RRc stars over 11 observing seasons.}
\label{fig:rrcpc}
\end{figure*}

Nearly all RRc stars show significant period changes over the
time-span of the analysed observations (11 yrs). It is well visible
in Fig.~\ref{fig:lc_rrc} as majority of the light curves cannot be
phased well with a single period. To study the period changes in
more detail, we used together data from both Swope and du Pont
telescopes. For each star and for both data sets we fixed the same
initial epoch and the same period (at which all data analysed
simultaneously phase best). Then we analysed the phase ($\phi_1$)
and amplitude ($A_1$) variation on the season-to-season basis. We
note that phase change diagrams are equivalent to O--C diagrams.
Only for V10, V15 and V23 phase changes are rather small and
insignificant. In other stars more complex patterns are observed
which we illustrate in Fig.~\ref{fig:rrcpc}. V33 is the only clear
case in which the overall phase change can be modelled with a
quadratic function. Such a phase change corresponds to a linear
period increase at a rate of $\dot{P}=0.71$\thinspace${\rm
d\,Myr^{-1}}$. This is two orders of magnitude faster than expected
due to the stellar evolution, see e.g. \cite{catelan}. We also note
that the fit for V33 is not perfect; residuals are significant, and
indicate possible more complex period changes on a shorter time
scale. In all other stars more complex phase changes are apparent,
examples are V11, V14 and V28 in Fig.~\ref{fig:rrcpc}. The phase
changes occur on a relatively short time-scale and are not
parabolic. Such phase changes are of non-evolutionary character and
are commonly observed in first overtone pulsators, both RR~Lyr stars
and classical Cepheids \citep[e.g.][]{szeidlM5,poleski}. Interesting
phase changes are detected for V21 and V22, as they might be
periodic, with periods of order of $\sim\!2600$\thinspace d. A
longer time-base of observations is needed for a confirmation,
however.

For all stars we also investigated the amplitude stability. In
contrast to the pulsation phases, the amplitudes are stable in most
cases (with the exception of Blazhko variables). Only in a few stars
the changes appear significant but the variation is below
$5$\thinspace per cent.

\begin{table*}
\centering
\caption{RRc stars with an additional shorter period variability in
         the $P_{\rm x}/P_{1}\!\in\!(0.60,\,0.65)$ range.
         Consecutive columns contain: star's id, first overtone
         period, $P_1$, additional period, $P_{\rm x}$, period
         ratio, $P_{\rm x}/P_1$, amplitude of the additional
         periodicity, $A_{\rm x}$, and indication which data were
         used in the analysis. Remarks in the last column: `c' --
         combination frequency, $\nu_1+\nu_{\rm x}$ detected, `a' --
         complex appearance of $\fSO$ signal (e.g. non-coherent,
         additional close peaks, broad power excess), `tdp' --
         time-dependent prewhitening used in the analysis, $P_1$
         determined from s1., `p' -- additional periodicity
         detected in the star; see Section~\ref{ssect:notesRRc} for
         details.}
\label{tab:61}
\begin{tabular}{lllllll}
id & $P_1$ (d) & $P_{\rm x}$ (d) & $P_{\rm x}/P_1$ & $A_{\rm x}$ (mag)  & data & remarks \\
\hline
V8  & 0.3814813   & 0.23975    & 0.6285 & 0.0026 & s1    & c (weak), p \\
V15 & 0.279945707 & 0.1715296  & 0.6127 & 0.0034 & all   & a        \\
V17 & 0.31460523  & 0.1924287  & 0.6117 & 0.0049 & s1-s4 &          \\
    &             & 0.1955796  & 0.6217 & 0.0045 &       &          \\
    &             & 0.1984814  & 0.6309 & 0.0045 &       &          \\
V21 & 0.281390043 & 0.1730099  & 0.6148 & 0.0033 & all   & a        \\
V23 & 0.275105063 & 0.1688913  & 0.6139 & 0.0023 & all   & c, a     \\
V24 & 0.3293664   & 0.2002600  & 0.6080 & 0.0026 & all   & tdp      \\
V27 & 0.2781221   & 0.1707740  & 0.6140 & 0.0024 & all   & a, tdp   \\
V33 & 0.3064223   & 0.187836   & 0.6130 & 0.0047 & s1    &          \\
    &             & 0.19332    & 0.6309 & 0.0037 &       &          \\
V35 & 0.29079074  & 0.1783040  & 0.6132 & 0.0029 & all   & a, p        \\
V36 & 0.310084    & 0.1897573  & 0.6120 & 0.0032 & all   & tdp      \\
\hline
\end{tabular}
\end{table*}

A study of period changes in RRab stars of NGC~6362 is more
difficult because of the strong Blazhko modulation in the majority
of these variables. Analysis for stars without the Blazhko
modulation shows no or negligible phase changes. For Blazhko
variables, periods must also be stable as in the majority of cases
we find no remnant power at $k\nu_0$ after prewhitening (see however
Section~\ref{ssec:notesab} and notes on V16 and V25). Also, a
comparison of light curves displayed in Figs. \ref{fig:lc_rrab} and
\ref{fig:lc_rrc} clearly shows, that in contrast to the RRc stars,
in the RRab stars periods are very stable -- in all cases data can
be phased well with a single period.

\subsection{Notes on individual RRc stars}\label{ssect:notesRRc}

\noindent{\it V8} -- the additional non-radial mode detected in this
star, reported in Tab.~\ref{tab:61}, is weak (${\rm S/N}=3.8$);
however, the presence of the combination frequency, $\nu_1+\nu_{\rm
x}$, supports the detection. In addition we detect a significant
(${\rm S/N}=5.5$) periodicity at $P_{\rm x2}=0.47427$\thinspace d
($A_{\rm x2}=5.6$\thinspace mmag), so $P_1/P_{\rm x2}=0.8043$. Such
a period ratio is expected for double-mode first and second overtone
pulsator, but only if the additional, weak periodicity corresponds
to the first overtone and the dominant variability to the second
overtone. We judge such possibility unlikely. Consequently,
additional long-period variability may be due to a g-mode
oscillation or to a contamination.

\smallskip

\noindent{\it V10} -- in the data prewhitened with the first
overtone frequency, its harmonics and all detected modulation side
peaks, we found an additional significant (${\rm S/N}=4.9$)
periodicity, $P_{\rm x}=1.03510$\thinspace d. This is most likely an
artefact, as we detect signals of nearly the same frequency in two
other stars, V25 and V11.

\smallskip

\noindent{\it V11} -- in the most numerous seasonal dataset, s1,
after prewhitening with the first overtone frequency and its
harmonics we detect an additional significant (${\rm S/N}=5.3$)
long-period variability, $P_{\rm x}=1.0363$\thinspace d. This
periodicity cannot correspond to any acoustic mode of oscillation.
This is most likely an artefact, as we detect signals of nearly the
same frequency in two other stars, V10 and V25.

\smallskip

\noindent{\it V35} -- in addition to a non-radial mode detected in
this star (Tab.~\ref{tab:61}) another long-period variability is
also present, $P_{\rm x2}=0.520838$\thinspace d ($A_{\rm
x2}=3.6$\thinspace mmag, ${\rm S/N}=4.8$). The period ratio is
$P_1/P_{\rm x2}=0.5583$. No combination frequencies are detected.
The additional variability cannot correspond to acoustic mode of
oscillation. It may be due to a g-mode oscillation or to a
contamination.

\smallskip

\noindent{\it V14, V22 and V28} -- except that in all cases period
changes are detected (which is well visible already in
Fig.~\ref{fig:lc_rrc}) we find no signature of modulation or of
additional modes in these stars.

\subsection{RRd stars}\label{ssect:rrd}

\begin{figure}
\includegraphics[width=\columnwidth]{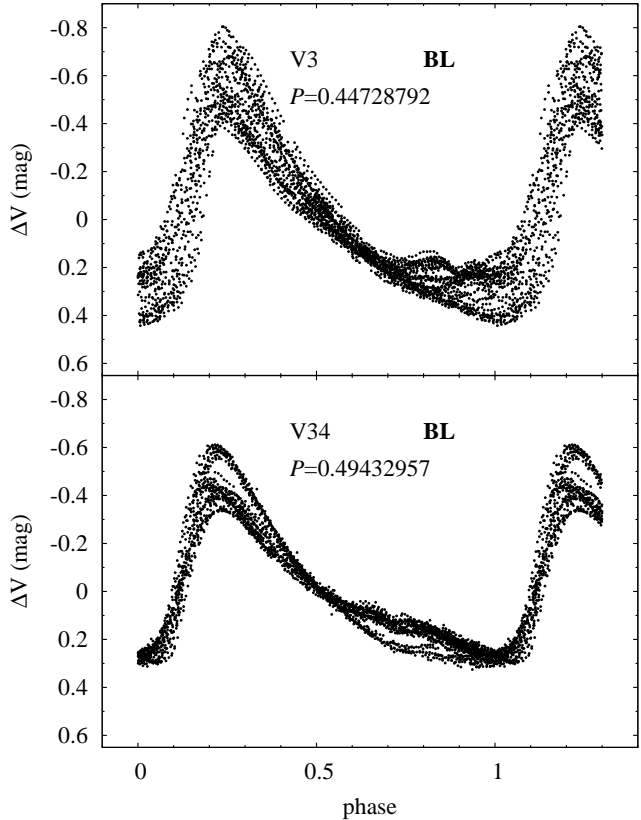}
\caption{Light curves of the two RRd variables phased with the
         fundamental mode period. In both cases the first overtone
         is of a low amplitude, and scatter present in both light
         curves is mostly due to Blazhko modulation.}
\label{fig:lc_rrd}
\end{figure}

\begin{figure}
\includegraphics[width=\columnwidth]{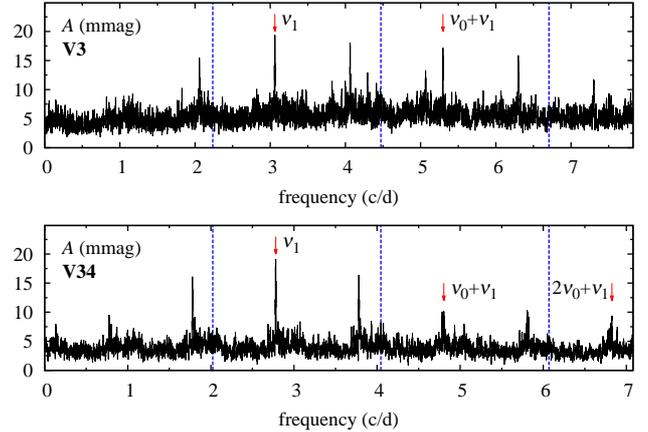}
\caption{Detection of the first overtone in V3 (top) and V34
         (bottom). The plots show the frequency spectrum for the two
         stars after prewhitening with the fundamental mode
         frequency, its harmonics (dashed lines) and all detected
         modulation peaks located close to $k\fF$. The pronounced
         signal from the first overtone is marked with an arrow.
         Several combination frequencies are detected and also
         marked.}
\label{fig:rrdfsp}
\end{figure}

Two stars previously identified as of RRab type, V3 and V34, are in
fact RRd pulsators. These are first RRd stars identified in
NGC~6362. Light curves phased with the fundamental mode period,
presented in Fig.~\ref{fig:lc_rrd}, show a significant scatter,
which is mostly due to the long-period modulation of the fundamental
mode, discussed later on in this section. In both stars, an
additional significant shorter-period variability is detected.
Fig.~\ref{fig:rrdfsp} illustrates the detection for both variables.
It shows frequency spectra after prewhitening with the fundamental
mode frequency, its harmonics and all significant modulation
components. Additional, very prominent, signals are marked with
arrows. Combination frequencies are detected as well, and are also
marked in Fig.~\ref{fig:rrdfsp} with arrows. Period ratios with the
fundamental mode period, $0.7304$ and $0.7275$, for V3 and V34,
respectively, suggest that the additional variability corresponds to
the radial first overtone. The period ratios are in fact lower than
typically observed in RRd stars of similar fundamental mode period.
V3 and V34 are new members of the {\it anomalous RRd} class recently
identified by \cite{anRRd_MC}. Other members of the class were
identified in the Galactic bulge, M3, and Magellanic Clouds
\citep[][respectively]{rs15a,jurcsikM3a,anRRd_MC}. V3 and V34 fit
well within the region occupied by these stars in the Petersen
diagram. Anomalous nature of the two RRd stars is discussed in more
detail in Section~\ref{ssect:anRRd}.

\begin{table*}
\centering
\caption{Properties of the two analysed RRd variables: star's id,
         fundamental mode and first overtone periods, period ratio,
         pulsation amplitudes of the fundamental and first overtone
         modes. Remarks: `BL' -- modulation of the radial mode(s)
         detected.}
\label{tab:d}
\begin{tabular}{lllllll}
id  & $P_0$ (d)  & $P_1$ (d) & $P_1/P_0$ & $A_0$ (mag) & $A_1$ (mag) & remarks \\ 
\hline
V3  & 0.44728792 & 0.3266817 & 0.7304 & 0.3771 & 0.0275 & BL \\
V34 & 0.49432939 & 0.3596333 & 0.7275 & 0.3089 & 0.0177 & BL \\
\hline
\end{tabular}
\end{table*}

\begin{table*}
\centering
\caption{Modulation properties of the anomalous RRd stars. Columns
         are the same as in Tab.~\ref{tab:bl} except the second
         column, which now identifies the pulsation mode. For each
         star, two rows are present: the first row refers to the
         modulation of the fundamental mode, the second row to the
         modulation of the first overtone.}
\label{tab:blrrd}
\begin{tabular}{llllrrrrlp{5.5cm}}
star id & mode & $P$ & $P_{\rm m}$ & $A_{-}$  & $A_{+}$ & $A_{\rm mod}$ & $Q$ & data & main modulation components\\
        &      & (d) &   (d)       & (mag)    & (mag)   &               &     &      & \\
\hline
V3      & F     & 0.44728792 & 310.07 & $0.0365$ & $0.0913$ & $0.242$ & $0.429$ & all & $\pm\nu_{\rm m1}$ components at $\nu_0,\ldots 4\nu_0$; $+\nu_{\rm m1}$ components at $5\nu_0,\ldots,7\nu_0$ \\
        & 1O    & 0.3266817  & 327.8  & $-$      & $0.0117$ & $0.425$ & $1.000$ &     & $+\nu_{\rm m2}$ component at $\nu_1$ \\
\noalign{\medskip}
V34     & F     & 0.49432939 & 784.3 & $-$      & $0.0421$ & $0.136$ & $1.000$ & all & $+\nu_{\rm m}$ components at $\nu_0,\ldots 7\nu_0$\\
        & 1O    & 0.3596333  & \multicolumn{7}{l}{no modulation detected} \\
\hline
\end{tabular}
\end{table*}

Periods and amplitudes of the two radial modes are given in
Tab.~\ref{tab:d}. In both cases, first overtone amplitude
constitutes only a small fraction of the fundamental mode amplitude,
$7.3$ and $5.7$ per cent, for V3 and V34, respectively. As already
mentioned, the fundamental mode is strongly modulated in both stars.
Properties of the modulation, determined from the analysis of the
frequency spectrum, are summarized in Tab.~\ref{tab:blrrd}.
Modulation periods, $\approx\!310$\thinspace d and
$\approx\!784$\thinspace d for V3 and V34, respectively, are
significantly longer than those in RRab or RRc stars. In both stars,
modulation side peaks at the fundamental mode frequency have higher
amplitude than the first overtone mode. Relative modulation
amplitudes of the fundamental mode are $\approx\!24$ and
$\approx\!14$ per cent for V3 and V34, respectively. In V3 a close
peak is also detected in the vicinity of $\nu_1$, indicating a
possible modulation of the first overtone with a period of
$\approx\!328$\thinspace d, which is slightly different from the
modulation period of the fundamental mode pulsation in the same star
($\approx\!310$\thinspace d; see Tab.~\ref{tab:blrrd}). The relative
modulation amplitude for the first overtone mode is large,
$\approx\!43$ per cent.

\subsection{The case of V37}\label{ssect:V37}

\begin{figure}
\includegraphics[width=\columnwidth]{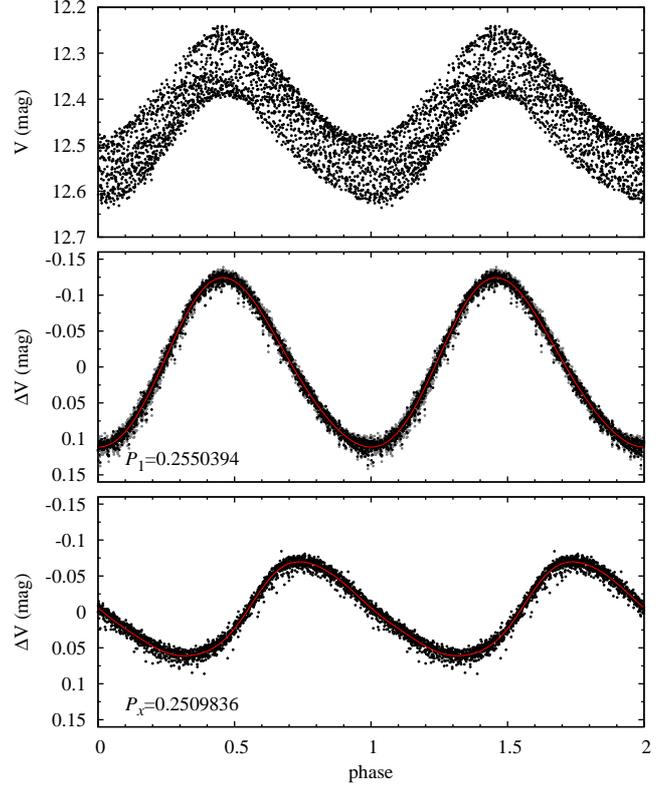}
\caption{Top panel: light curve of V37 phased with
         $P_1\! =\!0.2550394$\thinspace d. The large scatter is
         mostly due to beating of two periods, $P_1$, and a bit
         shorter period, $P_{\rm x}\! =\!0.2509836$\thinspace d.
         Disentangled light curves corresponding to these two
         periodicities are plotted in the middle ($P_1$) and in the
         bottom ($P_{\rm x}$) panels, with second and fourth order
         Fourier fits over-plotted (red solid line).}
\label{fig:37}
\end{figure}

V37 was identified by O01 as RR~Lyr star pulsating in the first
overtone. Its location in the period-amplitude diagram
(Fig.~\ref{fig:bailey}) and colour-magnitude diagram
(Fig.~\ref{fig:cmd}) is extreme, but based on these diagrams only,
the star could be regarded as hot, short-period RRc star of low
pulsation amplitude (likely located at the blue edge of the
instability strip). Its pulsation properties indicate it is not the
case, however. Its phased light curve, presented in the top panel of
Fig.~\ref{fig:37}, shows that the star is multiperiodic or
modulated. Indeed, the frequency spectrum of V37 is rich; three
independent periodicities are identified. Tab.~\ref{tab:37} provides
frequencies, amplitudes and phases of all significant peaks
identified in the frequency spectrum. The last two columns provide
two possible interpretations. In both of them, the frequency of the
highest amplitude peak is denoted by $\nu_1$; it corresponds to the
dominant period identified as due to radial first overtone in O01.
The second highest peak in the spectrum is located very close to
$\nu_1$  (its frequency is higher by $\approx\!0.0634$\thinspace
c/d) and can be interpreted twofold. In the first scenario
(`modulation?'; fourth column of Tab.~\ref{tab:37}), it is due to a
modulation of the dominant variability. The separation between the
two dominant peaks corresponds to the modulation frequency and is
denoted by $\nu_{\rm m1}$. Other significant peaks in the frequency
spectrum are components of the modulation multiplet, as given in the
fourth column of Tab.~\ref{tab:37}. In the second scenario
(`beating?'; fifth column of Tab.~\ref{tab:37}), the second highest
peak in the spectrum is treated as an independent frequency, denoted
by $\nu_{\rm x}$. Other peaks in the spectrum are then its harmonics
or linear combination frequencies with $\nu_1$. In addition to the
two dominant frequencies and their linear combinations, we also
detect an equidistant triplet centered on $\nu_1$ (peaks at
$\nu_1-\nu_{\rm m2}$ and at $\nu_1+\nu_{\rm m2}$), which, in both
scenarios, is attributed to modulation of $\nu_1$ with the
period of $\approx\!58.4$\thinspace d and with a rather low relative
modulation amplitude, of $5.7$\thinspace per cent. In the first
scenario, we have one dominant variability modulated with two
periods. In the second scenario, we deal with a double-periodic
variability, with a weak modulation of the dominant periodicity.
Which of the two scenarios is more plausible?

\begin{figure}
\includegraphics[width=\columnwidth]{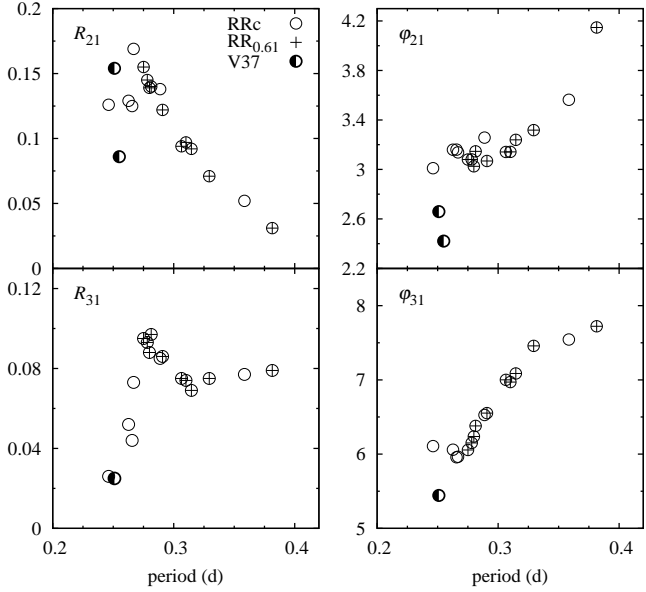}
\caption{Low-order Fourier decomposition parameters for RRc stars
         and for the peculiar variable V37 (half-filled symbols).
         For V37 two points are plotted for $R_{21}/\varphi_{21}$
         which correspond to two dominant periodicities detected in
         this star.}
\label{fig:fpc}
\end{figure}

A glimpse at the fourth column of Tab.~\ref{tab:37} points, that the
modulation scenario is unrealistic. Highly incomplete, high-order
multiplets are detected in the spectrum: quintuplet at $2\nu_1$,
septuplet at $3\nu_1$ and nonuplet at $4\nu_1$. Only the highest
frequency component of the nonuplet is detected in the vicinity of
$4\nu_1$ and only two highest frequency components of the septuplet
are detected in the vicinity of $3\nu_1$. The harmonics themselves,
$3\nu_1$ and $4\nu_1$, are not detected at all. Such a picture of
modulation is unrealistic, needles to say it was not detected in any
variable so far. At the same time, the beating scenario (fifth
column of Tab.~\ref{tab:37}) provides a simple and realistic
alternative explanation. Two close, highest amplitude peaks in the
frequency spectrum are interpreted as two independent periodicities.
One harmonic of $\nu_1$ and three harmonics of $\nu_{\rm x}$ are
detected, as well as several low-order combination frequencies which
are expected for double-periodic variability. In addition, the
dominant variability is modulated, as the equidistant triplet
centred on $\nu_1$ indicates. Period ratio, $P_{\rm x}/P_1=0.9841$,
indicates that the two periodicities cannot correspond to two radial
modes; at least one must correspond to a non-radial pulsation or its
origin is not due to pulsation at all. The two periodicities can be
easily separated and the resulting light curves are presented in the
middle and bottom panels of Fig.~\ref{fig:37}. Interestingly, it is
the lower amplitude light curve that is more non-sinusoidal, whereas
the light curve corresponding to the higher amplitude variability is
nearly symmetric. In both cases the corresponding Fourier parameters
are not typical for RRc variability -- see Fig.~\ref{fig:fpc}. In
particular, Fourier phase, $\varphi_{21}$, is significantly lower
than expected for RRc stars of similar period. Light curve shape,
position in the period-amplitude and colour-magnitude diagrams and
presence of two close, high-amplitude periodicities -- altogether
rule out the possibility that V37 is an RRc star. What kind of
variable is it, then?

The characteristic shape of the light curve plotted in the bottom
panel of Fig.~\ref{fig:37} ($P_{\rm x}$) suggests, it is due to
pulsation, most likely in the radial fundamental mode. In fact, the
light curve shape and the corresponding Fourier decomposition
parameters are typical for high amplitude $\delta$ Scuti stars
\citep[HADS; see eg.][]{poretti01}. These stars often pulsate in two
radial modes, fundamental and first overtone. The longest period of
the fundamental mode in these double mode variables is
$\approx\!0.22$\thinspace d \citep{poretti05}, but in other HADS it
can reach up to $\approx 0.3$\thinspace d. The period of V37 is a
bit longer than 0.22\thinspace d, and it pulsates in one radial mode
at most. However, genuine $\delta$~Sct stars are not expected to be
observed in NGC~6362, as the turn off mass in the cluster is around
0.8 solar masses \citep{jka15}. V37 could be a merger, just as it is
suspected for SX~Phe stars, six of which are detected in NGC~6362
\citep{jka14}. Light curves of SX~Phe stars can be similar to that
of V37, but their periods are much shorter and they are by $\approx
2$\thinspace mag fainter \citep{mazur99,jka14}.

The presence of the additional periodicity does not provide much
help; in fact its origin is even more puzzling. A non-radial mode of
a much higher amplitude and only slightly non-sinusoidal light curve
seems not probable. Slight asymmetry and short period,
$P_1\approx\!0.255$\thinspace d, seem to rule out the possibility
that this variability is due to rotation or binarity effects. We
stress that both periodicities are intrinsic to V37 as we detect
several strong combination frequencies, which are also prominent in
the flux data. Combination frequencies in the flux data, in
principle may also result from a non-linearity of the CCD detector.
However, two combination frequencies are detected also in the data
gathered with the du Pont telescope (these data are less numerous
and of inferior quality, though). Hence, we conclude that the two
dominant periodicities observed in V37 must be both intrinsic to
this variable. The puzzling frequency spectrum of V37 does not
result from blending of two stars.

The possibility that V37 is not a genuine member of NGC~6362 would
allow more speculations. We note however, that  \cite{zlo} lists V37
among genuine proper motion members of the cluster. Hence, location
of V37 in the colour-magnitude diagram, in particular, its  mean
brightness, typical for horizontal branch stars, is important
constraint for any explanation of its nature. Such explanation is
clearly lacking at the moment. We note that just recently, we have
discovered a very similar variable among RRc stars in the OGLE
Galactic bulge collection (Netzel \& Smolec, in prep.).
Its properties are also best explained as beating of two very close
periodicities. The light curve shapes are very similar to that
displayed in Fig.~\ref{fig:37} for V37. Hence, V37 might not be an
isolated, peculiar case, but may represent a new type of variability.

\section{Discussion}\label{sect:discussion}

\subsection{Blazhko effect in NGC~6362}\label{ssect:blazhko}

The incidence rate of the Blazhko phenomenon in NGC~6362,
$69$\thinspace per cent for RRab stars and $19$\thinspace per cent
for RRc stars, is very high -- among the highest reported in
ground-based observations published so far. We note however that
studies of the Blazhko modulation in globular clusters, based on
top-quality observations and Fourier analysis (which is essential
for firm confirmation of the Blazhko effect\footnote{Some studies
classify RR Lyr stars as Blazhko, based on the appearance of the
light curve only -- its scatter or non-repeating shape
\citep[e.g.][]{af12,kunder13}. With such an approach, the presence
of a quasi-periodic modulation cannot be firmly established. As a
result, spurious, exceedingly high incidence rates are sometimes
reported.}) are scarce. For M3, \cite{js16} reports $47$ and
$11$\thinspace per cent incidence rates for RRab and RRc stars,
respectively. In M5, $36$\thinspace per cent of RRab stars show the
Blazhko modulation \citep{jurcsikM5}. For the field, the top-quality
ground-based observations and space observations show, that the
incidence rate among RRab stars is close to $50$\thinspace per cent
\citep[e.g.][]{jurcsikKBS,geza}.

\begin{table}
\centering
\caption{Two interpretations for V37. The first three columns list
         frequencies, amplitudes and phases of significant peaks
         detected in the frequency spectrum of V37. The following
         two columns provide two possible interpretations:
         `modulation' and `beating' scenarios.}
\label{tab:37}
\begin{tabular}{r@{.}lllll}
\multicolumn{2}{c}{$\nu$\thinspace (c/d)} &  $A$\thinspace (mag)     & $\phi$\thinspace (rad)   & modulation?           & beating?\\
\hline
0&06336058 & 0.0053 & 5.82 & $\nu_{\rm m1}$          & $\nu_{\rm x}-\nu_1$ \\
3&90384907 & 0.0067 & 5.18 & $\nu_1-\nu_{\rm m2}$    & $\nu_1-\nu_{\rm m2}$ \\
3&92096263 & 0.1169 & 1.73 & $\nu_1$                 & $\nu_1$ \\
3&93807620 & 0.0029 & 2.31 & $\nu_1+\nu_{\rm m2}$    & $\nu_1+\nu_{\rm m2}$ \\
3&98432322 & 0.0645 & 6.09 & $\nu_1+\nu_{\rm m1}$    & $\nu_{\rm x}$ \\
4&04768380 & 0.0059 & 0.30 & $\nu_1+2\nu_{\rm m1}$   & $2\nu_{\rm x}-\nu_1$ \\
7&84192527 & 0.0101 & 5.88 & $2\nu_1$                & $2\nu_1$ \\
7&90528585 & 0.0070 & 3.60 & $2\nu_1+\nu_{\rm m1}$   & $\nu_{\rm x}+\nu_1$ \\
7&96864643 & 0.0099 & 2.26 & $2\nu_1+2\nu_{\rm m1}$  & $2\nu_{\rm x}$ \\
11&88960907 & 0.0043 & 5.71 & $3\nu_1+2\nu_{\rm m1}$ & $2\nu_{\rm x}+\nu_1$ \\
11&95296965 & 0.0017 & 4.86 & $3\nu_1+3\nu_{\rm m1}$ & $3\nu_{\rm x}$ \\
15&93729287 & 0.0012 & 1.33 & $4\nu_1+4\nu_{\rm m1}$ & $4\nu_{\rm x}$ \\
\hline
\end{tabular}
\end{table}

We stress that the reported high incidence rate of the Blazhko
phenomenon results not only from the top quality of the data we
analyse and consequently the low detection threshold in the mmag
range. In fact, modulation is easily detected because of large
modulation amplitudes. The amplitude of the dominant modulation side
peak in the vicinity of the radial mode frequency is at least
$3.5$\thinspace mmag for RRc stars and at least $9$\thinspace mmag
for RRab stars. Although the detection of $3.5$\thinspace mmag
signals is indeed challenging in typical ground based data,
identifying signals at the level of $9$\thinspace mmag is rather
routine in the data of massive sky surveys \citep[see e.g. analysis
of OGLE-III data by][]{netzel1}. The relative modulation amplitudes
for RRab stars range from $3$ to $24$\thinspace per cent and for RRc
stars from $2$ to $28$\thinspace per cent.

All modulation periods are below $100$\thinspace d with the
exception of the second modulation period detected in V30, which is
$\approx\!216.4$\thinspace d. V30 is the only RRab star in which two
modulation periods were detected. Two modulation periods were also
detected in RRc variable, V6, but the secondary modulation is very
weak. \cite{benko14} noted, based on the analysis of the {\it
Kepler} sample, that the ratio between the primary and secondary
modulation periods is nearly always close to a small integer
number. We note that the two modulation periods in V30 and V6 are
not in a resonant or close-to-resonant ratio as $P_{\rm m1}/P_{\rm
m2}$ is $\approx\!0.16$ for V30 and $\approx\!1.12$ for V6.

In the colour-magnitude diagram (Fig.~\ref{fig:cmd}) we marked the
Blazhko variables with thick symbols. Two of the three modulated RRc
stars, V6 and V10, are among the hottest RRc stars of NGC~6362. V36
on the other hand, in which the Blazhko modulation is weak (only one
modulation side peak was detected), is the coolest RRc star in the
sample. The majority of RRab stars show the Blazhko effect; they are
scattered over the entire domain of RRab stars in the diagram.

The only two RRd variables we identified in NGC~6362 show modulation
of the radial modes, which is characteristic for anomalous RRd
stars. Incidence rate of the Blazhko effect among M3 double-mode
variables is 45 per cent; 4 modulated RRd stars of M3 also have
anomalous period ratios \citep{jurcsikM3a,js16}; see next section
for more details.

\begin{figure*}
\includegraphics[width=2\columnwidth]{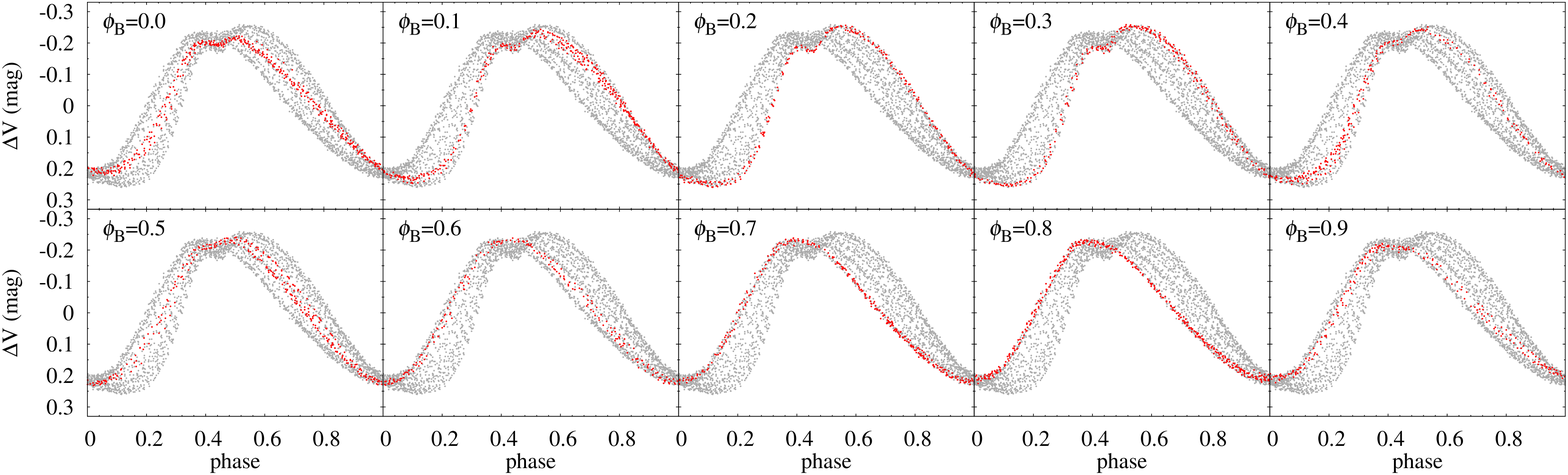}
\caption{Illustration of the Blazhko effect in RRc variable V10;
         snapshots from the animation available as supplementary
         on-line material.}
\label{fig:animV10}
\end{figure*}

Finally, we note that in some of the Blazhko variables, the phase
coverage of the modulation cycle is very good, and the Blazhko
effect itself appears regular. It allows to visualise the light
curve changes over the Blazhko cycle in the form of an animation.
Such animations for V5, V31 (RRab) and V10 (RRc) are available as a
supplementary on-line material. Data for these stars were phased
with the modulation period, divided into ten phase bins, and phased
with the pulsation period within each bin. Fig.~\ref{fig:animV10}
illustrates the light curve changes in V10. Both amplitude and phase
modulation are apparent. At phases of lower pulsation amplitude
($\phi_{\rm B}=0.6,\ldots,0.9$) the light curve is essentially
featureless. As pulsation amplitude gets higher, the shoulder (or
bump) appears on the ascending branch, just before the maximum
brightness.

\subsection{Anomalous RRd stars in NGC~6362}\label{ssect:anRRd}

\begin{figure}
\includegraphics[width=\columnwidth]{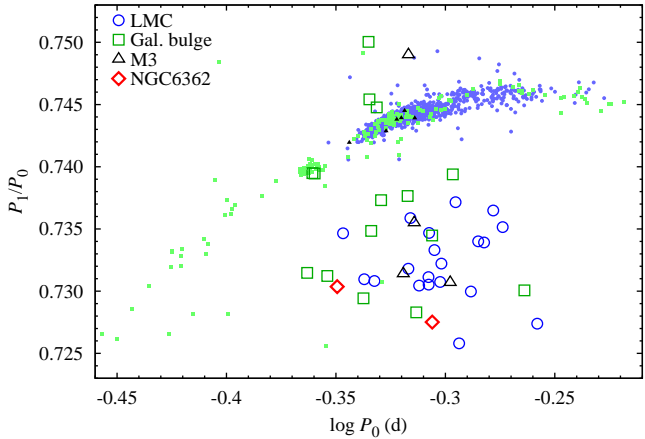}
\caption{Petersen diagram for classical and anomalous RRd stars of
         various stellar systems: Galactic bulge (green squares),
         LMC (blue circles), M3 (black triangles) and NGC~6362 (red
         diamonds). Classical and anomalous RRd stars are plotted
         with small filled symbols and with large open symbols,
         respectively.}
\label{fig:petrrd}
\end{figure}

Recently, the new class of multiperiodic RR~Lyr pulsators was
identified: anomalous RRd stars \citep{anRRd_MC}. In these
variables, two radial modes, fundamental and first overtone, are
involved in the pulsation, but otherwise properties of these stars
are anomalous, as compared to ``classical'' RRd stars. The first
objects of this type were discovered in the Galactic bulge
\citep{ogleiv_rrl_blg, rs15a} and in the globular cluster M3
\citep{jurcsikM3a}. Only recently additional objects were detected
in the OGLE Magellanic Clouds photometry by \cite{anRRd_MC}, who
proposed to call these stars anomalous RRd, based on their distinct
characteristics. Location of these stars in the Petersen diagram
does not follow the well defined progression delineated by
``classical'' RRd stars; their period ratios are significantly
different than in the classical RRd stars at the same fundamental
mode period. Typically, the period ratios are smaller. This is well
illustrated in Fig.~\ref{fig:petrrd}. Contrary to ``classical'' RRd
stars it is the fundamental mode that most often dominates the
pulsation. Finally, a long-term modulation of pulsation of one, or
of both radial modes is often detected in anomalous RRd stars.
\cite{anRRd_MC} also noted a somewhat peculiar light curve shape of
the dominant fundamental mode variation.

The two RRd stars newly detected in NGC~6362, V3 and V34, clearly
belong the class of anomalous RRd stars, as they share the same
characteristics. Their period ratios are lower than expected at a
given fundamental mode period. In both stars it is the fundamental
mode that dominates the pulsation. Finally, a modulation of the
radial modes is firmly detected in both V3 and V34.

It is important to note that anomalous RRd stars are not a
homogeneous group. Although all reported stars share the same broad
characteristics (e.g. anomalous period ratios, modulation), some
systematic differences can be easily pointed out while analysing
stars of different stellar systems. Contrary to classical RRd stars,
the fundamental mode usually has a larger amplitude than the first
overtone. In the Magellanic Clouds it is the case for 19 out of 22
stars, which is 86 per cent of the sample \citep[see tab.~1
in][]{anRRd_MC}. In M3, in three out of four anomalous RRd stars the
fundamental mode dominates \citep[$75$ per cent;][]{jurcsikM3a}.
Interestingly, the only star in which the first overtone dominates
the pulsation (V99) has a higher period ratio than the classical RRd
stars. In the Galactic bulge, only in 7 out of 15 stars reported by
\cite{rs15a} the fundamental mode dominates, which constitutes $47$
per cent of the sample. Still, this number is higher than in the
population of classical RRd stars in the same stellar system, where
it is only $18$ per cent. Further, if we exclude a few stars with
larger period ratios, in $70$ per cent of anomalous RRd stars of the
Galactic bulge the fundamental mode dominates. In the only two
anomalous RRd stars of NGC~6362, the fundamental mode strongly
dominates. In fact, the amplitude ratios, $7.3$ and $6.6$ per cent
in V3 and V34, respectively, are the smallest reported for anomalous
RRd stars.

Further differences are noticed in the modulation properties of RRd
stars with anomalous period ratios, most notably in the asymmetry of
modulation side peaks. In the modulated stars of the Magellanic
Clouds in all but one star, the modulation side peaks on the lower
frequency side of the radial mode are higher \citep[see tab.~2
in][]{anRRd_MC}. Relevant information is also available for the
Galactic bulge stars, in which there is no clear tendency: side
peaks on either side of the radial mode frequency can be higher
\citep[see tab.~2 in][]{rs15a}. In the two RRd stars of NGC~6362 the
modulation side peaks on higher frequency side of the radial mode
dominate or are the only side peaks detected.

A study of systematic differences between anomalous RRd stars of
different stellar systems is important for testing the models
explaining the origin of this peculiar form of pulsation. So far,
only one mechanism was proposed. Based on linear pulsation models,
\cite{anRRd_MC} pointed out that anomalous RRd pulsation might be
related to the parametric resonance of the form:
$2\nu_{1}=\nu_{0}+\nu_{2}$. In this scenario, a low damping rate of
the fundamental mode is needed to explain its typically higher
amplitude, as fundamental mode is a {\it daughter mode}, gaining
energy at the cost of the first overtone. A modulation of pulsation
amplitude and phase is also, in principle, possible in this
resonance scenario. Nevertheless, to validate the proposed
mechanism, and to provide any predictions that could be compared
with observations, non-linear model computations are needed.

\subsection{Non-radial pulsation in RRc stars} \label{ssect:nr}

\noindent {\it Earlier detections.} Detection of non-radial modes
was claimed in the observations of globular cluster M55
\citep{olech99}, and in NGC~6362 (O01). O01 lists three stars as
showing non-radial pulsations: V6, V10 and V37. V37 is very peculiar
object and we discussed it in detail in Section~\ref{ssect:V37}. Now
we focus on V6 and V10. The suspected non-radial modes were
identified by O01 based on the frequency spectrum analysis, as peaks
close to the radial mode frequency. In V6 and V10 two close peaks,
symmetrically placed around the radial mode frequency, were
detected. Such peaks, forming equidistant triplet with the radial
mode frequency, might also appear as due to modulation of pulsation,
i.e. the Blazhko effect. For these stars, the corresponding
modulation periods would be $15.5$\thinspace d and $8.5$\thinspace
d, for V6 and V10, respectively. Because at that time the known
Blazhko RRc stars were scarce and all Blazhko RRc stars reported in
the literature had longer modulation periods, O01 interpreted the
additional periodicities as due to non-radial pulsation. We also
note that some of the early models do explain the Blazhko effect as
a beating between the radial mode and non-radial, $\ell=1$ modes,
excited by the 1:1 resonance with the radial mode. The non-radial
modes are split due to star's rotation and form a symmetric triplet
with the radial mode frequency \citep{nd01}. The better and better
photometric data gathered over the past several years seem to
falsify such a model, though \citep[for recent reviews, see
eg.,][]{geza,sm16bl}. In many stars the modulation is clearly
not regular, consecutive modulation cycles, their length and shape,
differ. In several stars, including the ones reported here,
double-periodic modulation was detected. These observations cannot
be explained by clock-work models, like that of \cite{nd01}. In
addition, the close peaks seen in the frequency spectrum appear to
be parts of multiplet structures which are a signature of genuine
modulation. Due to the poor quality of the observations only one or
two components of the multiplet were detected in the past. Several
examples can be given for NGC~6362. When results of O01 and of the
present analysis are compared, we commonly detect triplets (V7, V18,
V30, V31, V32), full quintuplets or components of quintuplet
structure (V1, V5, V29, V6, V10) or even septuplet components (V13)
in place of a single close peak detected in the earlier study of
O01. This concerns both RRab and RRc stars. In particular, for V6
and V10 we find full quintuplet around at least one harmonic of
$\nu_1$, see the last column of Tab.~\ref{tab:bl}. These stars are
classified as genuine Blazhko RRc variables in our study. We also
note that several RRc stars with an even shorter modulation periods
are now known \citep[see e.g.][]{skarka}.
\smallskip

\begin{figure}
\includegraphics[width=\columnwidth]{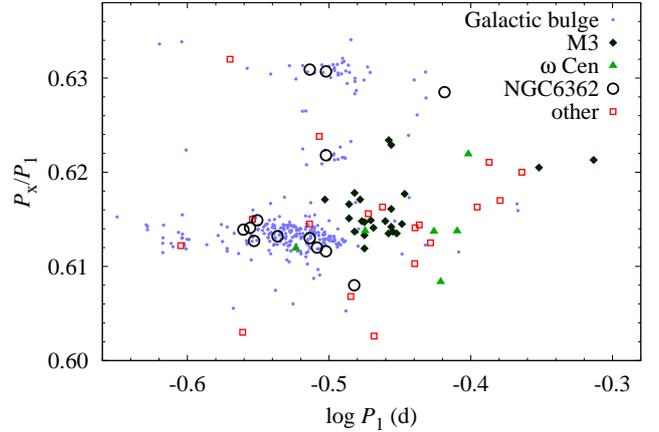}
\caption{Petersen diagram for RRc stars with additional variability
         in the $P_{\rm x}/P_{1}\!\in\!(0.60,\,0.65)$ range.}
\label{fig:pet61}
\end{figure}

{\it Variability in the $(0.60,\,0.65)P_1$ range.} Starting with
AQ~Leo observed with the space telescope {\it MOST} \citep{aqleo},
in many RRc and RRd stars additional signals in the $P_{\rm
x}/P_1\!\in\!(0.60,\,0.65)$ range and with amplitudes at the mmag
level were reported \citep[$\fSO$ peaks;
e.g.][]{om09,netzel1,netzel3}. This is particularly true for the
space observations, as nearly all RRc and RRd stars observed from
space show this form of pulsation
\citep[e.g.][]{szabo_corot,pamsm15,molnar}. Consequently, $\fSO$
frequencies must be common in the RRc/RRd variables, but it is
challenging to detect them from the ground, because of a higher
noise in the ground-based photometric data. In the Petersen diagram,
the stars group in three sequences: in the most populated, bottom
sequence, centred at $P_{\rm x}/P_1\!\approx\!0.613$, in the middle,
least populated sequence, centred at $0.623$, and in the top
sequence centred at $0.631$ (see Fig.~\ref{fig:pet61}). The middle
sequence was first recognized after analysis of the top-quality
sample of RRc stars observed by the OGLE project \citep{netzel3}. In
fact, most of the known \RRSO stars were discovered in the OGLE
Galactic bulge data and these stars are marked in
Fig.~\ref{fig:pet61} separately, with small bluish squares. The
above quoted average period ratios are from \cite{netzel3} and are
based on the OGLE Galactic bulge data only. In Fig.~\ref{fig:pet61}
we also included M3 sample from \cite{jurcsikM3b} (black diamonds),
$\omega$~Cen sample from \cite{om09} (green triangles) and all other
known stars of this type, mostly from space observations (red open
squares). In several stars, subharmonics of the additional
frequencies, i.e. peaks centred at ${1\over 2}\fSO$ and ${3\over
2}\fSO$ were reported \citep[see eg.,][]{pamsm15,molnar,kurtz}. In the
frequency spectrum, the $\fSO$ peaks and peaks at the subharmonic
frequencies are often non-coherent and appear as a clump of peaks or
as a wide power excesses.

The model of \cite{wd16} explains the ${1 \over 2}\fSO$ subharmonics
as due to non-radial strongly-trapped modes of $\ell=8$ (for the top
sequence in the Petersen diagram) and $\ell=9$ (for the bottom
sequence). Such modes have low observed amplitudes because of a
strong geometric cancellation and are rarely detected in \RRSO
stars. On the other hand, harmonics of these non-radial modes should
have higher observed amplitudes because of only weak geometric
cancellation and because of non-linear effects. These harmonics
correspond to the $\fSO$ periodicities commonly observed in RRc and
RRd variables. The middle sequence in the Petersen diagram of \RRSO
stars is explained by a combination frequency, $\nu_8+\nu_9$. In
this scenario, $\fSO$ peaks are in fact multiplets, which, due to
nonlinear mode interaction, have a complex form in the frequency
spectrum.

We note that $\fSO$ peaks are also detected in the first overtone
Cepheids \citep[e.g.][]{mk09,ogle_cep_smc}. The distribution of
period ratios of stars showing the $\fSO$ peaks with and without
subharmonics, both in Cepheids and in RR~Lyr stars, strongly
supports Dziembowski's model \citep{ss16}.

We do not detect subharmonics in \RRSO stars of NGC~6362. In one
star, V17, we detect periodicities that correspond to the three
sequences in the Petersen diagram (see Tab.~\ref{tab:61} and
Fig.~\ref{fig:61freq}). The `middle periodicity' should correspond
to the combination frequency then. In Fig.~\ref{fig:61freq}, we
indicate were the combination frequency is expected (arrow marked
with `c' label), based on the frequencies of the highest peaks
corresponding to the bottom and top sequences (marked with left-most
and right-most arrows in the same panel; $2\nu_8$ and $2\nu_9$
according to the Dziembowski's model). It indeed falls very close to
the location of the highest peak corresponding to the middle
sequence in the Petersen diagram. Although the agreement is not
perfect, it is satisfactory, taking into account the complex
appearance of the $\fSO$ peaks and consequently, intrinsic
difficulty in precise determination of their frequency. We note that
all other known RRc/RRd stars with periodicities corresponding to
the three sequences in the Petersen diagram pass this test
(Dziembowski \& Smolec, in prep.).
\smallskip

{\it Properties of \RRSO stars in the globular clusters.} A large
fraction of RRc stars in the globular cluster M3, $38$\thinspace per
cent (14 out of 37 RRc), are \RRSO stars. In addition, in 4 out 10
RRd stars ($40$\thinspace per cent) $\fSO$ peaks were detected
\citep{jurcsikM3b}. One of these RRd stars is anomalous. The
incidence rate is even higher among RRc stars of NGC~6362 as in 10
out of 16 RRc stars ($63$\thinspace per cent) we detect $\fSO$
peaks. It is the highest incidence rate in the published
ground-based observations. In the Petersen diagram,
Fig.~\ref{fig:pet61}, NGC~6362 variables nicely fit the three
sequences formed by the previously discovered objects, majority of
which are Galactic bulge stars. The only exception is V8 for which
period ratio is low, $0.6080$ (Tab.~\ref{tab:61}); still a few
objects with similar period ratios were reported in the literature
and are included in Fig.~\ref{fig:pet61}. In all but two stars we
detect only a single additional periodicity that falls within the
bottom sequence in the Petersen diagram. In V33 two additional
periodicities are detected, and they correspond to the bottom and to
the top sequence in the Petersen diagram. In V17 three periodicities
are detected that nicely fit to the three sequences delineated by
the OGLE Galactic bulge stars. A systematic difference with M3 stars
is apparent. First, \RRSO stars of M3 are shifted towards longer
periods. The shift is consistent with the difference of average RRc
periods in both clusters, which is $\langle P_{\rm RRc}\rangle_{\rm
M3}-\langle P_{\rm RRc}\rangle_{\rm
NGC6362}\!\approx\!0.042$\thinspace d. The scatter among \RRSO stars
of M3 seems larger; their mean period ratio is also slightly higher
than that derived from the Galactic bulge sample. In fact, the
Galactic bulge \RRSO stars are scarce at first overtone periods
characteristic for M3 \RRSO stars. Only six \RRSO stars were
identified in $\omega$~Cen \citep{om09}. Four of them have first
overtone periods longer than typically observed for \RRSO stars of
M3 and of NGC~6362. Two have shorter periods, one within the range
typical for M3 stars, and the other within a range typical for
NGC~6362 stars. In general, \RRSO stars of $\omega$~Cen are strongly
scattered over the Petersen diagram.

The described differences in the distribution of \RRSO stars of the
three globular clusters in the Petersen diagram must reflect the
metallicity and population differences between the three clusters.
Both M3 and NGC~6362 are OoI, but their mean metallicities differ,
${\rm [Fe/H]}=-1.57$ and ${\rm [Fe/H]}=-0.95$, for M3 and NGC~6362,
respectively. Also, M3 has been assigned to a young, while NGC~6362
to an old halo population \citep{ohyh} \citep[although the isochrone
fitting yields the same age for both systems][]{dotter}. The
metallicity difference alone cannot explain the observed period
shift between \RRSO stars of the two clusters. $\omega$~Cen, on the
other hand, is a very large and atypical globular cluster. It is
classified as OoII type and its mean metallicity (${\rm
[Fe/H]}=-1.62$) is similar to M3, but it is known to host diverse
stellar populations and might be a remnant of the dwarf galaxy
\citep[e.g.,][]{catelan}. It explains why its \RRSO stars are
scattered over the Petersen diagram; long periods of its four \RRSO
stars are consistent with its OoII designation.

Based on their analysis, \cite{jurcsikM3b} pointed out two
interesting properties of \RRSO variables, which we can check for
NGC~6362 stars. First, they noted, the \RRSO stars are located in
the colour-magnitude diagram (see their fig.~3) in the region
directly adjacent to the RRd domain (to its blue side; we note that
$\fSO$ peaks were also detected in RRd stars in their study). In the
colour magnitude-diagram plotted in our Fig.~\ref{fig:cmd}, we
marked the \RRSO stars with the plus sign. It is hard to define the
RRd domain in this plot, as we have only two anomalous RRd stars in
our sample. However, the interface between RRab and RRc stars is
clearly marked. No doubt, the \RRSO stars concentrate on the blue
side of this interface. We thus confirm the observation made by
\cite{jurcsikM3b}. Second, analysing the light curves and
corresponding Fourier decomposition parameters for RRc stars with
and without $\fSO$ peaks, \cite{jurcsikM3b} concluded (see their
fig.~4), that light curves of \RRSO stars are ``sinusoidal, with a
reduced, if any, bump preceding maximum brightness''. In
Fig.~\ref{fig:fpc} we plot the low-order Fourier decomposition
parameters for RRc stars and mark the \RRSO stars with plus sign.
Contrary to \cite{jurcsikM3b}, we observe that there is no
difference between $R_{21}$ and $R_{31}$ of \RRSO stars and other
RRc stars. Also, a glimpse at Fig.~\ref{fig:lc_rrc} indicates that
the bump preceding maximum brightness is commonly present in \RRSO
stars of NGC~6362 (e.g. in V23, V27, V15, V21). We conclude that
there is no specific feature of the light curve that distinguishes
the \RRSO phenomenon in general.

We do not detect subharmonics in \RRSO stars of NGC~6362.
Subharmonics were not detected in M3 either. \cite{jurcsikM3b}
detected $\fSO$ peaks in 4 RRd variables, including one anomalous
star. We find no trace of $\fSO$ peaks in the only two, anomalous
RRd stars of NGC~6362.

\smallskip

{\it Gravity modes in RRc stars?} In two RRc variables we have
detected additional long-period variability; too slow to be
associated with an acoustic mode of oscillation. On the other hand,
additional periods are too short to be connected with rotation or
with binarity effects. If these periodicities are intrinsic to the
stars, they would correspond to non-radial, gravity mode pulsation.
Detection of long-period additional variability in the RRc stars was
also reported in the analysis of top-quality {\it Kepler}
observations, see section~6 in \cite{pamsm15}. We stress that
possible detection of gravity modes in giant-type stars represents a
challenge to stellar pulsation theory \citep{wd77}.

In the case of V8 and V35 (see Section~\ref{ssect:notesRRc}) the
additional variability is in the mmag regime. In no case we have
detected combination frequencies with the dominant radial
oscillation, so we have no proof that additional periodicities are
intrinsic to these stars. Contamination is likely in the dense
fields of globular cluster. Consequently, we refrain from any
definite statements on the nature of additional long-period
variabilities in the discussed stars.

\section{Summary and conclusions}

We have analysed top-quality photometric data gathered for 35 RR~Lyr
stars in the globular cluster NGC~6362 within the CASE project. Our
most important findings are the following.

\begin{itemize}
\item Sixteen of the analysed stars are genuine RRab stars and other 16
      are genuine RRc stars, in agreement with earlier studies of
      RR~Lyr stars in NGC~6362.
\item Two stars previously classified as RRab, V3 and V34, are anomalous
      RRd stars, members of the recently identified class of
      double-mode pulsators. In addition to the radial fundamental
      mode we detect the radial first overtone of low amplitude.
      Period ratio of the two radial modes is significantly lower
      than for the majority of `classical' RRd stars at the same
      fundamental mode period. In both stars, either both radial
      modes are modulated (V3), or only one of the modes is
      modulated. Previously, anomalous RRd stars were identified in
      the Galactic bulge, the Magellanic Clouds and in the globular
      cluster M3.
\item V37, previously classified as RRc, shows beating of two close and
      large amplitude periodicities. The presence of combination
      frequencies in the flux data confirms that both are intrinsic
      to the star. Surprisingly, lower amplitude periodicity is more
      nonlinear; its characteristic shape, with short ascending
      branch and long descending branch, suggest that it is due to
      pulsation in the radial fundamental mode. The light curve
      corresponding to the dominant variability is more symmetric; a
      weak modulation of this signal was also detected. Both light
      curves significantly differ from the shape typical for RRc
      stars. Its RR~Lyr classification is thus tentative; however,
      we cannot propose other satisfactory explanation of its
      nature. We note that the star is a genuine proper-motion
      member of the cluster and its brightness is typical for
      horizontal branch stars in NGC~6362.
\item Ten out of 16 RRc stars show additional periodicities in the
      $(0.60,\,0.65)P_1$ range. These double-periodic variables,
      abbreviated here as RR$_{0.61}$, are well known; in fact,
      based on space observations, it is expected that this form of
      pulsation must be common among RRc/RRd stars. The incidence
      rate in NGC~6362, $63$\thinspace per cent, is the highest in
      ground-based observations published so far. Properties of
      \RRSO stars in NGC~6362 are qualitatively similar to
      properties of \RRSO stars in the Galactic bulge; in particular
      their location in the Petersen diagram is very similar. \RRSO
      stars of NGC~6362 fall within the three sequences formed by
      the Galactic bulge stars. Systematic differences are noted
      when \RRSO stars of NGC~6362 are compared with \RRSO stars of
      other globular clusters, M3 and $\omega$~Centauri. Numerous
      populations of \RRSO stars in different stellar systems are
      essential for testing the models which try to explain the
      nature of \RRSO stars.
\item Our results support the observation made by \cite{jurcsikM3b} in M3:
      \RRSO stars tend to cluster on the cool side of the RRc
      domain, at the interface with RRab domain, where RRd pulsation
      is also expected.
\item Incidence rate of the Blazhko effect is among the highest reported
      in the ground-based observations. We detect modulation in 11
      out of 16 RRab stars ($69$\thinspace per cent) and in 3 out of
      16 RRc stars ($19$\thinspace per cent). In addition, one RRc
      star and one RRab star are rare examples of stars in which two
      modulation periods are detected.
\item RRc stars are prone to phase instabilities. These rarely correspond
      to steady period increase/decrease, but in the majority of
      cases have complex nonlinear appearance and occur on time
      scales much shorter than expected due to evolution. In
      contrast, majority of RRab stars have stable pulsation periods
      over the time span of the observations.
\end{itemize}

\section*{Acknowledgements}
PM and RS were supported by the grants DEC-2012/05/B/ST9/03932 and
DEC-2015/17/B/ST9/03421, and JK and WP were supported by the grant
DEC-2012/05/B/ST9/03931 from the National Science Centre, Poland.







\section*{Supporting information}
Additional Supporting Information may be found in the online version of this article:
\smallskip

\noindent{\bf AnimationV5} -- Animation shows light curve changes over the Blazhko cycle for V5.
\smallskip

\noindent{\bf AnimationV10} -- Animation shows light curve changes over the Blazhko cycle for V10.
\smallskip

\noindent{\bf AnimationV31} -- Animation shows light curve changes over the Blazhko cycle for V31.



\bsp    
\label{lastpage}
\end{document}